\documentclass[preprint,rmp,12pt]{revtex4}
\usepackage{epsf}
\usepackage{times}
\usepackage{bm}
\usepackage{big}
\begin{document}

\def\tr{{\mathrm T}}
\def\rd{{\mathrm d}}
\def\e{{\mathrm e}}
\def\i{{\mathrm i}}
\def\ssum{{\mbox{\small{$\Sigma$}}}}
\newcommand{\mb}[1]{\ensuremath{\bm{#1}}}

\author{Magnus Rattray and Jonathan L. Shapiro}

\affiliation{\mbox{Computer Science Department,}  \\ 
\mbox{University of Manchester,} \\
\mbox{Manchester M13 9PL,} \\ 
\mbox{United Kingdom.}\\ \\ \\ \\ \\ \\ \\ \\ \\
\mbox{Corresponding author:}\\
\mbox{Magnus Rattray,} \\
\mbox{Department of Computer Science,} \\
\mbox{University of Manchester,} \\
\mbox{Manchester M13 9PL, UK.} \\ \\
\mbox{Running title: Cumulant dynamics}}

\date{\today\newpage}

\title{Cumulant dynamics of a population under multiplicative
selection, mutation and drift}

\begin{abstract}

We revisit the classical population genetics model of a population
evolving under multiplicative selection, mutation and drift. The
number of beneficial alleles in a multi-locus system can be considered
a trait under exponential selection. Equations of motion are derived
for the cumulants of the trait distribution in the diffusion limit and
under the assumption of linkage equilibrium. Because of the additive
nature of cumulants, this reduces to the problem of determining
equations of motion for the expected allele distribution cumulants at
each locus. The cumulant equations form an infinite dimensional linear
system and in an authored appendix Adam Pr{\"u}gel-Bennett provides
a closed form expression for these equations. We derive approximate
solutions which are shown to describe
the dynamics well for a broad range of parameters.  In particular, we
introduce two approximate analytical solutions: (1) Perturbation
theory is used to solve the dynamics for weak selection and arbitrary
mutation rate. The resulting expansion for the system's eigenvalues
reduces to the known diffusion theory results for the limiting cases
with either mutation or selection absent. (2) For low mutation rates
we observe a separation of time-scales between the slowest mode and
the rest which allows us to develop an approximate analytical solution
for the dominant slow mode. The solution is consistent with the
perturbation theory result and provides a good approximation for much
stronger selection intensities.

\end{abstract}

\maketitle

\section{Introduction}

When modelling the dynamics of a finite population it is necessary to
consider the effects of fluctuations introduced by genetic drift. The
most influential theoretical approach to this problem has been to
formulate a model of the allele frequency dynamics which incorporates
drift and selection at the genic level (for models of molecular
evolution we might be considering individual nucleotides but we will
adopt gene nomenclature and talk of the alternative alleles at each
locus). The resulting model can be analysed using the theory of Markov
chains or more usually by invoking a diffusion approximation~\citep[see,
for example,][]{ewens,crow}. Alternatively, a number of theoretical
models of selection have been developed which describe evolution at the
phenotype level. These models typically describe the population by a
small number of descriptive statistics, such as moments or cumulants of
a trait distribution. Such models are useful when describing selection
on quantitative traits~\citep{bulmer,tb,dawson,burger91} and have also been used to
develop theories of asexual evolution in finite
populations~\citep{higgs,adam97}. \citet{burger91} provided one of the
first analyses of polygenic dynamics using cumulants and in a recent
book he provides a detailed account of the cumulant dynamics
associated with various evolutionary
models~\citep{burger00}. Cumulants have also been used for modelling
the population dynamics of genetic algorithms, where an accurate
characterisation of finite population effects is crucial~\citep[see,
for example,][]{jls94,rattray96}.  Cumulants have a number of features
which make them attractive for modelling polygenic
dynamics. Cumulants of increasing order will often have decreasing
impact, a feature not shared by the distribution moments. Another
important feature is that cumulants are additive, in the sense that
the cumulants of a sum of random variables is equal to the sum of the
cumulants of each random variable. Standard cumulants are most
appropriate when describing deviations from a Gaussian trait
distribution~\citep{kendall} while factorial cumulants can be useful
for distributions close to Poisson~\citep{dawson}.

In this work we explore the relationship between the allele frequency
dynamics and trait distribution dynamics using a diffusion
approximation.  Dynamical equations for the trait distribution
cumulants are derived for a finite population undergoing
multiplicative selection and reversible mutation, assuming
recombination maintains linkage equilibrium. Because of the additive
nature of cumulants, this reduces to the problem of determining
equations of motion for the expected allele distribution cumulants at
each locus. This is a classical model in population genetics for which
the non-equilibrium diffusion theory is not solved for the general
case, although solutions exist for the restricted cases when either
mutation or selection is absent~\citep{crow}. Analysis of the cumulant
dynamics allows a new perspective as well as some new results. The
cumulants form an infinite dimensional linear dynamical system and two
approximate analytical solutions are found which accurately describe
the dynamics for a broad range of parameter values. In the first case
we use perturbation theory to solve the dynamics for weak selection
and arbitrary mutation rate. The perturbation expansion is consistent
with Kimura's eigenvalue expansion for the zero mutation
case~\citep{kimura55}. Secondly, we show how a separation of
time-scales between the leading mode and the rest can be used to
develop an approximation to the dynamics which holds for low mutation
rates and a greater range of selection intensities. This separation of
time-scales allows a remarkably simple and novel approximation to the
transient dynamics under selection and weak mutation. The theory is
compared to simulation results of finite populations undergoing free
recombination, showing good agreement.

\section{Equations of motion}

\subsection{The model}

We consider a haploid population which is assumed to be at linkage
equilibrium. The population has two alleles $x_i\in\{1,0\}$ at each of
$L$ loci distributed according to the following factorised
distribution,
\begin{equation}
	\phi({\bm x}) = \prod_{i=1}^L \left( p_i\delta_{x_i,1} +
	(1-p_i)\delta_{x_i,0}\right) \ ,
\label{eq:px}
\end{equation}
where $\phi({\bm x})$ represents the probability mass function and
$\delta_{x,y}$ is the Kronecker delta. Each population member is
labelled $n=1,2,\ldots,N$ where $N$ is the population size and we
consider multiplicative selection with the fitness defined,
\begin{equation}
	w_n = \prod_{i=1}^L (1-s)^{1-x_i^n} \ .
\end{equation}
We also include symmetric mutation with rate $u$ at each generation.

\subsection{Allele frequency dynamics}

If we take $s\sim O(N^{-1})$ and $u\sim O(N^{-1})$ then for large $N$
the allele frequency at each locus is subject to an independent
diffusion process. We identify $\alpha\equiv sN$ and $\beta\equiv uN$
to be the relevant dimensionless quantities of order unity. The
dynamics is completely determined by the mean and variance of the
allele frequency change each generation and under the standard
multinomial selection model we find~\citep[see, for
example,][]{ewens},
\begin{eqnarray}
	\mbox{E}(\Delta p_i|p_i) & = & a(p_i) \delta t\quad
	\mbox{where} \quad a(p_i) = \alpha p_i(1-p_i) + \beta (1-2p_i)
	\ , \nonumber \\ \mbox{Var}(\Delta p_i|p_i) & = & b(p_i)
	\delta t\quad \mbox{where} \quad b(p_i) = p_i(1-p_i) \ ,
\label{eq:diffusion}
\end{eqnarray}
to leading order in $\delta t$, where $\delta t\equiv N^{-1}$ defines
an increment of time. The diffusion limit is obtained as $\delta t
\rightarrow 0$. The covariance between allele frequencies at different
loci is $O(s^2\delta t)$ and is therefore negligible in this limit, so
we are justified in treating each locus independently as long as
recombination is sufficient to maintain linkage equilibrium. In the
diffusion framework the Kolmogorov forward equation (also known as the
Fokker-Planck equation) describes the evolution of the allele
frequency distribution over time,
\begin{equation}
	\frac{\partial \phi(p_i,t)}{\partial t} =
\frac{1}{2}\frac{\partial^2}{\partial
p_i^2}\left[b(p_i)\phi(p_i,t)\right] - \frac{\partial}{\partial
p_i}\left[a(p_i)\phi(p_i,t)\right] \ ,
\end{equation}
where $\phi(p_i,t)$ is the probability density of $p_i$ at time
$t$. This PDE is difficult to solve analytically and to our knowledge
no analytical solution has been determined for the general case of
selection, mutation and drift. In this case numerical methods have to
be used to study the dynamics of the allele frequency
distribution. The solution at equilibrium is easier to obtain and was
given by~\citet{kimura55} confirming the result originally found
by~\citet{wright} using a different approach. For a good discussion of
diffusion theory in the context of population genetics see, for
example,~\citet{crow}.

We have assumed that selection acts independently at each locus. This
assumption can break down if $L$ is sufficiently large, because
interference between loci results in a reduced effective population
size and the introduction of significant linkage disequilibrium.  This
effect can be quantified if the population is at linkage equilibrium
before selection by calculating the fitness variance in a finite
population. The effective population size can then be determined using
the arguments developed by~\citet{robertson}.  However, such
interference will also generate strong linkage
disequilibrium after selection which remains significant even under
free recombination.  We will therefore assume that $L$ is sufficiently
small to make such effects negligible.  A necessary condition for this
to be the case is for $s^2=o(L^{-1})$ while sufficient conditions
depend on the rate of recombination.

\subsection{Cumulants at linkage equilibrium}

Under multiplicative selection we can consider the number of
advantageous alleles per population member $Z_n\equiv\sum_{i=1}^L
x_i^n$ to be a trait undergoing exponential selection. The first two
cumulants are the mean and variance of the trait distribution while
higher cumulants describe deviations from a Gaussian distribution. Let
$K_m$ be the $m$th cumulant of the population. Each cumulant can be
generated by differentiating the appropriate generating
function~\citep{kendall},
\begin{equation}
	K_m = \lim_{\gamma\rightarrow 0}
	\frac{\partial^m}{\partial\gamma^m} G(\gamma) \quad
	\mbox{where} \quad G(\gamma) = \log\sum_{n=1}^N \e^{\gamma
	Z_n} \ .
\label{eq:kappa}
\end{equation}
If we assume a sufficiently large population at linkage equilibrium
then Eq.~(\ref{eq:kappa}) can be written in terms of allele
frequencies by replacing the sum in Eq.~(\ref{eq:kappa}) by an average
over the allele distribution defined in Eq.~(\ref{eq:px}),
\begin{eqnarray}
\label{def_Kn}
	G(\gamma) & = & \log \sum_{\bm x} \phi(\bm x) \e^{\gamma
	\sum_i x_i} \nonumber \\ & = & \sum_{i=1}^L
	\log\left(p_i(\e^\gamma-1)+1\right) \ .
\end{eqnarray}
In terms of allele frequencies the first few cumulants are,
\begin{eqnarray}
	K_1 & = & \sum_{i=1}^L p_i \ , \nonumber \\ K_2 & = &
	\sum_{i=1}^L p_i (1-p_i) \ , \nonumber \\ K_3 & = &
	\sum_{i=1}^L p_i (1-p_i) (1-2p_i) \ .
\label{eq:K123}
\end{eqnarray}
The first cumulant is the number of advantageous alleles within the
population and the second cumulant measures the heterozygosity. From
these equations we see that the expectation value of the trait
cumulants is linearly related to the allele frequency moments.

Notice that each cumulant can be written as a sum over contributions
from each locus. This is not generally true for other combinations of
the trait distribution's moments, such as the central moments. Because
each locus is effectively subject to an independent diffusion process
the central limit theorem ensures that fluctuations in each cumulant
over an ensemble of populations undergoing an independent diffusion
process will decrease with increased $L$. In the large $L$ limit the
cumulants will evolve deterministically and will be described by the
deterministic equations of motion derived below (recall, however, that
we require $s^2=o(L^{-1})$ for our approximation to be valid, which
limits the size of $L$). For finite $L$ our equations of motion will
only give the mean dynamical trajectory over an ensemble of
populations. Figure~\ref{fig:fluc} shows how fluctuations in the
cumulant dynamics are reduced as $L$ is increased. In
section~\ref{sec:fluc} we will show how the scale of these
fluctuations can be estimated from the expectation value of the
cumulants.

\subsection{Cumulant equations of motion}

To calculate how the generating function for the cumulants changes
over time we follow the discussion given
by~\citet[p~136]{ewens}. First we write,
\begin{equation}
	G(\gamma) = \sum_{i=1}^L g_\gamma(p_i) \quad \mbox{where}
	\quad g_\gamma(p_i) = \log\left(p_i(\e^\gamma-1)+1\right) \ .
\end{equation}
Taylor expanding in orders of $\Delta p_i$ and taking the expectation
it is straightforward to show that,
\begin{equation}
	\mbox{E}_{t+\delta t}[g_\gamma(p_i)] \simeq
	\mbox{E}_t[g_\gamma(p_i)]+\mbox{E}_t[a(p_i)g_\gamma'(p_i)+\frac{1}{2}b(p_i)g_\gamma''(p_i)]\delta
	t \ ,
\end{equation}
where $a(p_i)$ and $b(p_i)$ are defined in
Eq.~(\ref{eq:diffusion}). In the limit $\delta t \rightarrow 0$ we
therefore obtain the rate of change of the generating function for the
expectation value of the cumulants,
\begin{equation}
	\frac{\rd}{\rd t} \mbox{E}_t[G(\gamma)] =
	\mbox{E}_t\left[\sum_{i=1}^L
	a(p_i)g_\gamma'(p_i)+\frac{1}{2}b(p_i)g_\gamma''(p_i)\right] \
	.
\label{eq:dGdt}
\end{equation}
The rate of change in the expectation value for each cumulant is given
by differentiating Eq.~(\ref{eq:kappa}),
\begin{equation}
	\frac{\rd}{\rd t} \mbox{E}_t[K_n] = \lim_{\gamma\rightarrow 0}
	\frac{\partial^n}{\partial\gamma^n}
	\mbox{E}_t\left[\sum_{i=1}^L
	a(p_i)g_\gamma'(p_i)+\frac{1}{2}b(p_i)g_\gamma''(p_i)\right] \
	.
\label{eq:dKdt}
\end{equation}
In order to get equations of motion for the cumulants we first
calculate the right hand side of Eq.~(\ref{eq:dKdt}) in terms of
allele frequencies. We then use the relationship between the allele
frequencies and cumulants in Eq.~(\ref{def_Kn}) to obtain equations
involving only cumulants. We automated this process using the symbolic
programming language Mathematica~\citep{math}. However, since
submitting this article Pr\"{u}gel-Bennett has obtained a closed form
expression for the dynamical equations which is given in the authored
appendix~\ref{app_adam}.

We obtain a linear first order ODE for the expected cumulants
$k_n(t)\equiv \mbox{E}_t[L^{-1}K_n]$ (we scale the cumulants by a
factor of $L^{-1}$ for convenience),
\begin{equation}
	\label{dkdt} \frac{\rd {\bm k}}{\rd t} = -{\bm M}{\bm k} +
	{\bm d} \ ,
\label{eq:dkdt}
\end{equation}
where ${\bm d}=(\beta,\beta,\beta,\ldots)^T$ and,
\begin{equation}
	{\bm M}=\left(\begin{array}{c c c c c} 2\beta & -\alpha & 0 &
	0 & \cdots \\ 0 & 1+4\beta & -\alpha & 0 & \cdots \\ 2\beta &
	0 & 3(1+2\beta) & -\alpha & \cdots \\ 0 & (1+8\beta) & 0 &
	2(3+4\beta) & \cdots \\ \vdots & \vdots & \vdots & \vdots &
	\ddots \end{array}\right) \ .
\end{equation}
Unfortunately, the equation for each cumulant involves a cumulant of
higher order, so that the system is intrinsically infinite
dimensional. In the next section we describe some approximate
solutions to this system of equations.

The cumulant equations derived above can easily be related to a set of
equations describing the moments of the allele frequency distribution,
since the cumulants are linearly related to these moments by
Eq.~(\ref{eq:K123}). However, the results described in the next
section will often rely on special features displayed by the cumulants
which are not shared by the allele frequency moments. In particular,
the truncated system obtained by setting moments above some order to
zero is very poorly behaved in contrast to the truncated cumulant
system used in sections~\ref{sec:truncation} and~\ref{sec:adiabatic}
below.

\section{Solving the dynamics}

The solution to Eq.~(\ref{eq:dkdt}) can be written in terms of a
diagonal matrix of eigenvalues ${\bm D} = [\lambda_i\delta_{ij}]$ and
the associated matrix ${\bm V}$ whose columns are eigenvectors of
${\bm M}$ (defined by ${\bm D}={\bm{V^{-1} M V}}$),
\begin{equation}
\label{k_soln}
	{\bm k}(t) = {\bm k}^*+{\bm V}\,\e^{-{\bm D}t}\,{\bm
	V}^{-1}({\bm k}(0) - {\bm k}^*) \ .
\label{eq:k(t)}
\end{equation}
Here, ${\bm k}^\ast\equiv {\bm M}^{-1}\bm d$ is the fixed
point. Properties of the fixed point are described in
appendix~\ref{app:fp}.

We have not been able to find a general result for the eigenvalues and
eigenvectors of the full infinite dimensional system. In practice we
find that a truncated system describes the dynamics well for a large
range of parameters, as we will show in
section~\ref{sec:truncation}. In section~\ref{sec:pert} we use a
perturbation expansion to solve the infinite dimensional system for
small $\alpha$. This expansion agrees with the zero mutation rate
result due to~\cite{kimura55} but only provides a good approximation
for $\alpha\simeq 3$ or less. In section~\ref{sec:adiabatic} we give
an approximation valid for low mutation rates which describes the
system accurately for larger values of $\alpha$. This approximation is
based on the observation that there is a separation of time-scales
between the slowest mode and the rest as the mutation rate is
reduced. This slow mode is strongly coupled to the first cumulant and
determines its expected rate of change. The higher cumulants quickly
relax to a quasi-equilibrium which can be determined analytically for
a truncated system. The second cumulant (trait variance) of this
quasi-equilibrium distribution determines the rate of change in the
first cumulant, and hence the eigenvalue of the slow mode.  The
separation of time-scales persists for relatively large selection
intensities and comparisons with simulations suggest that the
resulting approximation works well up to at least $\alpha\simeq 10$.

\subsection{Truncated system}
\label{sec:truncation}

A finite system of equations can be obtained by truncating the
infinite system in Eq.~(\ref{eq:dkdt}). We create an $n$-dimensional
truncated system by only considering cumulants of order $n$ and
less. Setting $K_{n+1}$ to zero decouples the equations from higher
cumulants. The solution is given by Eq.~(\ref{eq:k(t)}) with the
infinite-dimensional matrix $\bm M$ replaced by the $n$-dimensional
square submatrix starting in its top left-hand corner. The resulting
finite system provides a very good approximation for the dynamics of
the lower order cumulants as long as $\alpha$ is not too large. An
increased order of truncation is required to achieve accurate results
as $\alpha$ is increased. This is demonstrated in Fig.~(\ref{fig:fp})
where we compare the fixed point of the cumulant equations for the
truncated system to the known exact result derived from Wright's
distribution (see appendix~\ref{app:fp}). As the truncation order
increases the approximation rapidly converges toward the correct
result. However, for larger $\alpha$ the speed of convergence is
slower. It is interesting to note that the results do not seem to
depend strongly on the mutation rate. This suggests that the truncated
system will work well for a large range of mutation rates but may
break down for large $\alpha$. In particular, we have limited our
simulations to $\alpha\leq10$ although a higher order truncation might
allow accurate results for stronger selection.

In Fig.~(\ref{fig:dynamics}) we compare the solution of the eighth
order truncated system to averaged simulation results. The symbols
show the first two cumulants for $\alpha=1$, $5$ and $10$ and solid
lines show the theory. We observe excellent agreement between the
theory and simulations. A similar truncated system describing the
dynamics of the allele frequency moments does not reproduce the
dynamics or fixed point well unless selection is very weak ($\alpha\ll
1$) or mutation rates are high.

\subsection{Perturbation theory}
\label{sec:pert}

Perturbation theory allows the solution to an intractable linear
system by expansion around a solvable limiting case~\citep[see, for
example,][]{bender}. For $\alpha=0$ the Jacobian matrix $\bm M$ is
lower triangular and the eigenvalues and eigenvectors can be
determined exactly even for the full infinite dimensional
system. Indeed, the diffusion theory for $\alpha=0$ has an exact
closed form solution~\citep{kimura64} and one can verify that the
known eigenvalues agree with our zeroth order result. We expand around
the zeroth order result to obtain a perturbation expansion of
arbitrary order, as described in appendix~\ref{app:perturbation}.

The 4th order expansion is given for two eigenvalues by
Eqs.~(\ref{eq:eval1}) and (\ref{eq:eval2}). For small $\beta$ these
are the smallest eigenvalues and we observe that $\lambda_1\propto
\beta$ for all orders of expansion. For $\beta=0$ the first eigenvalue
is therefore zero and $\lambda_2$ determines the rate of decay under
selection and drift. The expansion for $\lambda_2$ in this limiting
case agrees with the expansion found by~\citet{kimura55} from an
eigenvalue expansion of the diffusion equation. To our knowledge no
similar result has previously been obtained for the case of mutation,
selection and drift studied here.

In Fig.~\ref{fig:eval} we compare the $2$nd and $6$th order
perturbation expansion result to eigenvalues of a truncated
system. The truncation order was increased well beyond the point where
the eigenvalues were observed to converge onto a limiting value. These
figures suggest that the perturbation theory provides a good
approximation up until about $\alpha=3$, which agrees with Kimura's
observation in the zero mutation case~\citep{kimura55}.

\subsection{Separation of time-scales and adiabatic elimination}
\label{sec:adiabatic}

We will often be interested in the limiting case where the mutation
rate is small. In this limit we find there is a separation of
time-scales between the slowest mode, which is $O(\beta)$, and the
rest, which are $O(1)$. This picture was confirmed for small $\alpha$
by the perturbation theory described in section~\ref{sec:pert} and
appendix~\ref{app:perturbation}. Figure~\ref{fig:eval} suggests that
this behaviour also holds for larger selection intensities. In
practice the dynamics will quickly relax to the slow mode, which will
determine most of the system's long term
dynamics. Figure~\ref{fig:pop_vel} shows the averaged velocity of a
population as it approaches the fixed point. The simulation results
approach an asymptotic value (derived below) as $\beta$ is
decreased. Notice that the velocity is proportional to $\beta$ and
that the slow mode becomes increasingly dominant as $\beta$ is
reduced. Similar simulations were used by \citet{adam99} in order to
compare the dynamics of populations with and without crossover,
although the scaling relationship with the mutation rate was not
considered explicitly in that study.

Once the population reaches the steady state it moves as a travelling
wave, as has previously been observed by~\citet{burger93}, in marked
contrast to the infinite population dynamics. B{\" u}rger argues that
the velocity of the population is proportional to the population's
variance, which can be determined under the assumption of a balance
between neutral mutation and drift. In the present case mutation is
not neutral, and does change the population mean, so that B{\"
u}rger's approximation is not strictly applicable. Also, we find that
the population's variance depends on the selection intensity and
cannot be calculated under a neutral model. However, the picture of a
travelling wave is still applicable and for our model we can derive an
analytical expression for the population's velocity in the steady
state in a similar spirit to B{\" u}rger's approximation.

The asymptotic result shown in Fig.~\ref{fig:pop_vel} is obtained by
assuming an adiabatic elimination of fast variables for small
$\beta$~\citep[for a discussion on adiabatic elimination see, for
example,][]{gardiner}. We make the assumption that most of the first
cumulant dynamics is determined by the slow mode. This is borne out by
examination of the eigenvectors derived from perturbation theory,
which show that the fast modes only couple to the first cumulant
through terms which are $O(\beta)$ (this is demonstrated in
appendix~\ref{app:adiabatic}). The higher cumulants have significant
contributions from the fast modes. We can therefore think of the
higher cumulants as fast variables which will rapidly converge to
a quasi-equilibrium value which then changes slowly according to the
slow mode.  The coupling of the higher cumulants to the slow mode is
through the first cumulant, so the quasi-equilibrium value is
determined by solving the fixed point equations for the second and
higher order cumulants for a particular value of the first
cumulant. To calculate this fixed point we have to truncate the
dynamical equations, as described in section~\ref{sec:truncation}, and
the quasi-equilibrium cumulants are found by solving a linear
equation,
\begin{equation}
	\sum_{j=2}^n M_{ij}k_j^{\mbox{\tiny qe}} = \beta - M_{i1}k_1
	\quad \mbox{for} \quad 2\leq i\leq n \ ,
\end{equation}
where $n$ is the truncation order. We then use the quasi-equilibrium
variance $k_2^{\mbox{\tiny qe}}$ to determine the rate of change of
the first cumulant according to Eq.~(\ref{eq:dkdt}),
\begin{equation}
	\frac{\rd k_1}{\rd t} = \alpha k_2^{\mbox{\tiny
	qe}}+\beta(1-2k_1) \ .
\label{eq:dk1dt}
\end{equation}
For an eighth-order cumulant truncation we find the following
expression for the quasi-equilibrium variance to first order in
$\beta$,
\begin{equation}
	k_2^{\mbox{\tiny qe}} \simeq \beta \left(1 +
	\frac{126\,\alpha\,(2k_1-1)(4200+220\alpha^2+\alpha^4)}{1587600
	+ 189000\alpha^2 + 2898\alpha^4 + \alpha^6}\right) \ .
\label{eq:k2qe}
\end{equation}
Higher order truncations result in similar rational expressions with
higher order polynomials in $\alpha$ appearing in the numerator and
denominator. The above expression was used to plot the theoretical
lines in Fig.~(\ref{fig:pop_vel}). The adiabatic elimination result is
self-consistent since Eq.~(\ref{eq:dk1dt}) demonstrates that only the
$O(\beta)$ leading eigenvalue contributes to the decay of the first
cumulant, as we initially assumed.

For weak selection, perturbation theory confirms the adiabatic
elimination result. The term proportional to $k_1$ on the right hand
side of Eq.~(\ref{eq:dk1dt}) gives the leading eigenvalue and we can
expand in $\alpha$,
\begin{equation}
	\lambda_1 \simeq 2\beta\left(1 + \frac{\alpha^2}{3} -
	\frac{\alpha^4}{45} + \cdots \right) \ .
\end{equation}
Comparison with Eq.~(\ref{eq:eval1}) to first order in $\beta$ shows
that the results agree. However, the simulations in
Fig.~\ref{fig:pop_vel} indicate that the adiabatic elimination result
provides a good approximation for much larger selection intensities
than covered by the perturbation theory.

\subsection{Fluctuations}
\label{sec:fluc}

We have mainly considered the averaged cumulant dynamics. However, the
theory described here can also be used to estimate fluctuations
from expected behaviour if we assume the same initial conditions at
all loci. In this case there is an identical and independent diffusion
process at all loci and we can consider
the expected cumulants to be ensemble averages over loci and/or
populations. For example, the variance of the mean is then given by,
\begin{eqnarray}
\mbox{Var}(L^{-1}K_1) & = &
		\mbox{E}_t\left[\left(\frac{1}{L}\sum_{i=1}^L
		p_i\right)^2\right]-\mbox{E}_t\left[\frac{1}{L}\sum_{i=1}^L
		p_i\right]^2 \nonumber \\ & = & L^{-1}[k_1(1-k_1)-k_2]
		\ .
\end{eqnarray}
where $k_n\equiv \mbox{E}_t[L^{-1}K_n]$. This shows how the
fluctuations fall off with increasing $L$, as previously observed in
Fig.~\ref{fig:fluc}. We can similarly calculate the ensemble variance
for the higher cumulants. It should be noted that the cumulants here
describe an infinite population from which our finite population can
be considered a sample. This sampling procedure will also introduce
fluctuations in the measured cumulants, as quantified
by~\citet{adam97}.

\section{Conclusion}

We studied the cumulant dynamics for a classical population genetics
model, a finite population with two alleles at each locus evolving
under multiplicative selection and reversible mutation. We identified
the number of advantageous alleles as a trait undergoing exponential
selection and the cumulants of the trait distribution were shown to be
linearly related to moments of the allele frequency distribution. The
dynamical equations were derived and formed a coupled
infinite-dimensional linear system. A truncated system was shown to
provide an excellent approximation to the full infinite-dimensional
system for a broad range of parameters. For weak selection we
developed an analytical approximation to the eigensystem of the full
infinite-dimensional case using a perturbation expansion around the
solvable limit of zero selection. To all orders in perturbation theory
a single eigenvalue was shown to be proportional to the mutation rate,
resulting in a separation of time-scales between different modes for
low mutation rates. This separation of time-scales was observed to
persist for stronger selection and allowed us to develop an analytical
approximation to the leading eigenvalue in the limit of weak
mutation. This approximation was shown to provide good results for
larger selection intensities than covered by the perturbation
expansion.

Cumulants have been shown to be useful for modelling a variety of
polygenic systems~\citep[see, for example,][]{burger00} and here we
have shown their potential for describing a finite population where
genetic drift is non-negligible.  Cumulants have a number of features
which make them useful for modelling finite populations close to
linkage equilibrium.  Firstly, it was noted that each cumulants at
linkage equilibrium can be written as a sum over contributions from
each locus. This is not generally the case for other combinations of
the trait moments, such as the central moments, and suggests that
populations may be better described using cumulants rather than
central moments since fluctuations will be reduced according to the
central limit theorem. For the multiplicative selection case
considered here the equations of motion were also linear, in which
case fluctuations do not lead to strong systematic effects. This
picture should be contrasted with cumulant equations describing finite
asexual populations under multiplicative selection which are
intrinsically far from linkage equilibrium, in which case fluctuations
do contribute systematic effects and must be modelled
explicitly~\citep{adam97}. Secondly, although the trait cumulants are
linearly related to the allele frequency moments, the latter do not
share certain advantageous features displayed by the former. This is
because the truncated equations of motion for the allele frequency
moments do not describe the infinite-dimensional system well except
under weak selection or high mutation rates.

Whether cumulant dynamics will prove generally useful for describing
finite populations with significant linkage disequilibrium is an open
question. Infinite populations with linkage disequilibrium have
recently been modelled using factorial cumulants~\citep{dawson} but as
we mentioned above, drift will introduce systematic effects which are
typically difficult to model. Although approximations have been
introduced which attempt to capture these fluctuations~\citep{adam97},
the complexity and non-linearity of the resulting dynamical system
make it difficult to draw general conclusions.

\subsection*{Acknowledgements}

We would like to thank Adam Pr\mbox{\"u}gel-Bennett and Nick Barton
for useful comments on a preliminary version of this paper.

\appendix

\section{Fixed Point}
\label{app:fp}
The fixed point of the allele frequency distribution for the model
considered here was first determined
by~\citet{wright}. \citet{kimura55} later showed this to also be the
solution of the related diffusion equation with density given by,
\begin{equation}
	\phi(\{p_1,p_2,\ldots,p_L\}) = C^{-L}\prod_{i=1}^L \e^{2\alpha
	p_i}\left[p_i(1-p_i)\right]^{2\beta-1} \ ,
\end{equation}
where,
\[
	C=\frac{\Gamma(2\beta)^2
	F(2\beta,4\beta,2\alpha)}{\Gamma(4\beta)} \ ,
\]
and $F(;,;,;)$ is the Kummer confluent Hypergeometric function. The
moments of this distribution can be shown to be related through the
following recursion formula,
\begin{equation}
	2\alpha\mbox{E}(p_i^n) =
(2-n+2\alpha-4\beta)\mbox{E}(p_i^{n-1}) -
(2-n-2\beta)\mbox{E}(p_i^{n-2}) \ .
\end{equation}
We can use this to determine relationships between the expected
population cumulants using the definition in Eq.~(\ref{def_Kn}). For
the first few we find,
\begin{eqnarray}
\label{kwright}
	\alpha k_2 & = & \beta(2k_1-1)\nonumber \ , \\ \alpha k_3 & =
	& (1+\alpha+4\beta)k_2 - 2\beta k_1\nonumber \ , \\ \alpha k_4
	& = & 3(1-\alpha+2\beta)k_3 + (3+4\alpha+12\beta)k_2 - 6\beta
	k_1 \nonumber \ .
\end{eqnarray}
As a check one can easily verify that these relationships satisfy
Eq.~(\ref{eq:dkdt}) with $\rd\bm k/\rd t=0$. In an authored appendix
to this paper, Pr\"{u}gel-Bennett shows that the expressions are
exactly equivalent (appendix~\ref{app_adam}).

\section{Perturbation theory}
\label{app:perturbation}

Consider the linear system in Eq.~(\ref{eq:dkdt}). To determine the
properties of the dynamics, we should study the eigenvalues and
eigenvectors of \mb{M}. This we do below by expanding around the
solution for $\alpha=0$.

\subsection{Notation}
Denote the $i$th eigenvalue of $\bm{M}$ as $\lambda(i)$; its
associated eigenvector is $\bm{e}(i)$. Denote the associated
eigenvector of the transpose of $\bm{M}$ as $\bm{w}(i)$. In
Eq.~(\ref{eq:dkdt}), $\bm{V}$ is a matrix whose $i$th column is
$\bm{e}(i)$ and $\bm{V^{-1}}$ is a matrix whose $i$th row is
$\bm{w}(i)^\dagger$.

We will expand in powers of $\alpha$ and use superscripts to denote
the order. So, $\bm{M}$ is written as
\begin{equation}
  \label{eq:matrix_expand} \bm{M} = \bm{M}^0 - \alpha \bm{\Delta} \ ,
\end{equation}
where $\Delta_{ij} = \delta_{i j+1}$. We likewise expand the
eigenvalues and eigenvectors,
\begin{eqnarray}
  \label{eq:expansion} \bm{e(i)} &=& \bm{e}^0(i) + \alpha \bm{e}^1(i)
  + \alpha^2 \bm{e}^2(i) + \dots \ , \\ \lambda(i) &=& \lambda^0(i) +
  \alpha \lambda^1(i) + \alpha^2 \lambda^2(i) +\dots \ ,\\ \bm{V} &=&
  \bm{E}^0 + \alpha\bm{E}^1 + \alpha^2 \bm{E}^2 +\dots \ ,\\
  \bm{V^{-1}} &=& \bm{W}^0 + \alpha\bm{W}^1 + \alpha^2 \bm{W}^2 +\dots
  \ , \\ \bm{D} &=& \bm{D}^0 + \alpha \bm{D}^1 + \alpha^2 \bm{D}^2
  +\dots \ .
\end{eqnarray}
The expansion is simple due to two properties of $\bm{M}^0$,
\begin{enumerate}
\item $\bm{M}^0$ is lower triangular, and
\item $\bm{M}^0$ alternates with $0$'s. i.e. $M^0_{ij}\neq 0
\Rightarrow M^0_{i j+1} = M^0_{i+1 j} = 0$.
\end{enumerate}

\subsection{Zeroth Order}
\label{sec:0th}

Since $\bm{M}^0$ is lower triangular, the eigenvalues are simply the
diagonal elements,
\begin{equation}
  \label{eq:eigenvalues0} \lambda^0(k) = M_{kk} = \frac{k(k-1)}{2} +
  2k\beta \ .
\end{equation}
The eigenvectors can be found recursively via back-substitution,
\begin{equation}
  \label{eq:Reigenvector0} e^0(k)_j = \left\{ \begin{array}{ll} 0;&
  \mbox{if $j<k$}\\ 1;& \mbox{if $j=k$}\\ \sum_{m=k}^{j-1}\frac{
  M_{km} e^0(k)_m}{\lambda^0(k)- \lambda^0(j)};&\mbox{otherwise}
  \end{array}\right. \ .
\end{equation}
The eigenvectors of the transposed system can also be found. The
result is
\begin{equation}
  \label{eq:Leigenvector0} w^0(k)_j = \left\{ \begin{array}{ll}
  \sum_{m=j+1}^{k}\frac{ M_{jm} w^0(k)_m}{\lambda^0(k) -
  \lambda^0(j)};&\mbox{if $j<k$}\\ 1;& \mbox{if $j=k$}\\ 0;& \mbox{if
  $j>k$}\\ \end{array}\right. \ .
\end{equation}
The latter can be found either by truncating the transposed system at
some order $n$, computing the first $n$ eigenvectors by
back-substitution starting from the lower right, and then taking
$n\rightarrow\infty$, or by ansatz. We will argue momentarily that
these are the only eigenvectors of the infinite system.

It is clear that the right eigenvectors form a complete set and are
defined without requiring any finite truncation\footnote{They are not,
however, normalizable in infinite dimensions}. The left eigenvectors
can be found only by truncating the system to some order or by
assuming that $w^0(k)_j = 0$ for all $j > k$. Thus, one might worry
that there are other possible left eigenvalues. However, since all of
the eigenvalues are distinct, the right and left eigenvectors must
form a mutually orthogonal set. Those given in
Eqs.~(\ref{eq:Reigenvector0}) and (\ref{eq:Leigenvector0}) do, and so
the left eigenvectors must form a complete set as well.

The result is that $\bm{E}^0$ and $\bm{W}^0$ have the same structure
as $\bm{M}^0$; lower triangular with alternating zero elements.

\subsection{Arbitrary Order}
\label{sec:arborder}

The $n$th order approximation to the eigenvalues and eigenvectors can
be found in terms of the lower order approximations. The recursion is
%
%
%
\begin{equation}
  \label{eq:perturbn} \bm{M}^0 \bm{E}^n - \bm{E}^n \bm{D}^0 - \bm{D}^n
\bm{E}^0 = \Delta\bm{E}^{n-1} + \sum_{j=2}^{n-1} \bm{E}^{n-j}\bm{D}^j
\ .
\end{equation}
This must be solved for the $n$th order contribution to the
eigenvectors, $\bm{E}^n$, and to the eigenvalues, $\bm{D}^n$.  This is
solvable because $\bm{E}^0$ is lower triangular, so the upper
triangular part of $\bm{E}^n$ can be computed without knowledge of
$\bm{D}^n$. Then, since $\bm{M}^0 - \bm{D}^0$ is lower triangular with
zeros on the diagonal, $\bm{D}^n$ depends only on the upper triangular
part of $\bm{E}^n$. Once this is computed, the lower triangular part
of $\bm{E}^n$ can be found.

This equation does not determine the diagonal part of $\bm{E}^n$. This
is arbitrarily taken to be zero. In other words, $V_{ii} = 1$ to all
orders which is like a choice of normalization for these
non-normalizable vectors.

To find the left eigenvectors, we ensure that $\bm{W}$ is the inverse
of $\bm{E}$ to $n$th order in $\alpha$. This requires that
\begin{equation}
  \label{eq:perturbwn} \bm{W}^n = - \left(\sum_{j=0}^{n-1}\bm{W}^j
  \bm{E}^{n-j}\right)\bm{W}^0 \ .
\end{equation}

We find that each order introduces one higher upper diagonal element
to \mb{E} and \mb{W}. Thus, to compute $\lambda(k)$ to order $n$, one
needs to truncate the system at an order higher than $k+n$.

\subsection{Results}
\label{sec:results}
The expansion for the first two eigenvalues is given below,
\begin{eqnarray}
  \label{eq:eval1} \lambda(1)&=&2 \beta + \alpha^2 \frac{2 \beta}{3 +
  10 \beta + 8 \beta^2} - \alpha^4 \frac{2 \beta (3 - 6 \beta -
  4\beta^2 + 16 \beta^3)}{(1+2 \beta)^3 (3+4 \beta)^3 (5+ 4 \beta)} +
  O(\alpha^6) \ , \\ \label{eq:eval2} \lambda(2) &=& 1+ 4 \beta +
  \alpha^2\frac{1+8 \beta^2}{2 (1+\beta)(1 + 2 \beta)(5+4\beta)} \\ &
  & \mbox{}- \alpha^4 \frac{1 - 278 \beta + 492 \beta^2 + 1392\beta^3
  - 96 \beta^4 - 2112\beta^5 -640 \beta^6 + 512\beta^7}{8 (1+\beta)^3
  (1+ 2 \beta)^3 (5+4 \beta)^3(7+4 \beta)} + O(\alpha^6)\nonumber \ .
\end{eqnarray}
As seen in Fig.~\ref{fig:eval}, the series expansion to $6$th order in
$\alpha$ is accurate for $\alpha$ up to about $3$.

\section{Adiabatic elimination}
\label{app:adiabatic}

The dynamical equation can be written as
\begin{equation}
  \label{eq:perturbMore_dynamics} \bm{V}^{-1} \frac{d\bm{K}}{d t} =
  -\bm{D} \bm{V}^{-1}\bm{k} + \bm{V}^{-1} \bm{d} \ .
\end{equation}
To all orders in perturbation theory and to leading order in $\beta$,
this equation takes the following form,
\begin{eqnarray}
  \label{eq:perturbMore_expansion} \dot{k_1} + \alpha f_{1 2}
  \dot{k_2} + \alpha^2 f_{1 3} \dot{k_3} + \dots &=& e^{-\lambda(1)t}
  \left( k_1 + \alpha f_{1 2}k_2 + \alpha^2 f_{1 3}k_3 + \dots\right)
  + \mbox{ constant},\nonumber\\ \beta \alpha f_{2 1}\dot{k_1} +
  \dot{k_2} + \alpha f_{2 3} \dot{k_3} + \dots &=& e^{-\lambda(2)t}
  \left(\beta \alpha f_{2 1} k_1 + k_2 + \alpha f_{2 3}k_3 +
  \dots\right) + \mbox{ constant},\nonumber\\ \beta f_{3 1}\dot{k_1} +
  \alpha f_{3 2}\dot{k_2} + \dot{k_3} + \dots &=& e^{-\lambda(3)t}
  \left(\beta f_{3 1} k_1 + \alpha f_{3 2} k_2 + \alpha k_3 +
  \dots\right) + \mbox{ constant},\nonumber\\ \vdots & &\vdots
\end{eqnarray}
where $\dot{k_n}$ is the time-derivative of $k_n$ and the $f_{ij}$'s
are trivially related to the eigenvectors but have the leading order
in $\beta$ pulled out.  The structure of the eigenvectors and
eigenvalues reveals three points,
\begin{itemize}
\item For all equations beyond the first one, coupling to $k_1$ and
$\dot{k_1}$ is through terms which are $O(\beta)$.
\item The first eigenvalue is $O(\beta)$; the remaining ones are
$O(1)$.
\item Truncating the equations to have $m$ components results in an
error of $O(\alpha^{m-i+1})$ to the $i$th equation.
\end{itemize}
This shows why adiabatic elimination works. For $\beta = 0$,
$(k_2,k_3.\ldots)$ satisfies a closed set of equations completely
decoupled from $k_1$. Once the higher cumulants have decayed, $k_1$
becomes frozen. For $\beta$ small, the coupling is present but weak
and the higher cumulants will behave as a quasi-closed set decaying at
a rate of $O(1)$. The first cumulant will decay very slowly after the
higher ones have decayed and will depend on them through their
quasi-equilibrium values. As long as the selection intensity is not
too large, truncating the system will result in small errors.

\section{Explicit Form of the Cumulant Evolution Equations}
\label{app_adam}

\newcommand{\Kcoef}[2]{{\Bigl[ {#1 \atop #2} \Bigr]}}
\newcommand{\Mcoef}[2]{{\Bigl\{ {#1 \atop #2} \Bigr\}}}
\newcommand{\Kcoefs}[2]{{\bigl[ {\scriptstyle #1 \atop #2} \bigr]}}
\newcommand{\Mcoefs}[2]{{\bigl\{ {\scriptstyle #1 \atop #2} \bigr\}}}
\newcommand{\binom}[2]{{\Bigl( {#1 \atop #2} \Bigr)}}
\newcommand{\pred}[1]{\mathrm{I}\!\left[ #1 \right]}
\renewcommand{\e}[1]{\mathrm{e}^{#1}}
\newcommand{\av}[1]{\mathrm{E_t}\!\left[ #1 \right]}
\newcommand{\dd}{\mathrm{d}} \newcommand{\ii}{\mathrm{i}}

\noindent{\it Adam Pr\"ugel-Bennett\\ ECS, University of Southampton,
SO17 1BJ, United Kingdom }
\vspace{2ex}

In this appendix we give an explicit form for the evolution
equation~(12).  We also obtain an analogous equation for the moments.
We can rederive the diffusion equation for the allele frequency
distribution (equation~(4)) directly from the matrix equation for the
moments.  This illustrates clearly the connection between a matrix
equation formulation and a diffusion equation formulation.

This work was completed after receiving a preprint of the above paper
by Rattray and Shapiro.  It seems appropriate to publish these results
at the same time as the paper.

\subsection{Explicit Evolution Equation}

Equation~(12) can be written explicitly in terms of binomial
coefficients
\begin{equation}
\frac{\dd k_n}{\dd t} = \beta + \alpha k_{n+1} - \sum_{m=2}^{n}
\left(2\beta \, \binom{n}{m-1} + \binom{n}{m-2} \right)
\pred{\mbox{$n+m$ is even}} \, k_m \label{eq:evol}
\end{equation}
where we denote the indicator function by
\begin{displaymath}
\pred{\mbox{\textit{predicate}}} = \left\{
\begin{array}{lll}
1 & \hspace{1em} & \mbox{if \textit{predicate} is true} \\ 0 &
\hspace{1em} & \mbox{if \textit{predicate} is false}
\end{array} \right. 
\end{displaymath}

We can write an equivalent set of equations for the allele frequency
moments $\mu_n=\av{p^n}$,
\begin{equation}
  \label{eq:moments-diff} \frac{\dd \mu_{k+1}}{\dd t} = -(k+1) \left(
  \alpha \mu_{k+2} + \left(2\beta-\alpha+\frac{k}{2}\right) \mu_{k+1}
  - \left(\beta+\frac{k}{2}\right) \mu_k \right).
\end{equation}
This equation still holds for $k=0$ provided we interpret $\mu_0=1$.
We can obtain the diffusion equation~(4) directly from this equation.

\subsubsection*{Sketch of Proof}

The proof of equation~(\ref{eq:evol}) follows from two observation.
Unfortunately, it involves rather laboured algebraic manipulations.
We therefore give only an outline of the proof below.

The first observation (already discussed in section II) is that the
cumulant equations all decouple over the loci.  We can therefore
consider the cumulants for a single locus.  We can write an analogous
equation to equation~(11) for the single-locus cumulants (we drop the
loci subscript $i$ as it plays no part in what follows)
\begin{equation}
  \label{eq:sl-dynamics} \frac{\dd k_n}{\dd t} = \av{ a(p)
  \frac{\partial K^{sl}_n(p)}{\partial p} + \frac{b(p)}{2}
  \frac{\partial^2 K^{sl}_n(p)}{\partial p^2} }
\end{equation}
with $a(p)=\alpha\,p\,(1-p) + \beta (1-2p)$, $b(p)=p\,(1-p)$,
\begin{equation}
  \label{eq:Ksl} K^{sl}_n(p) = \frac{\partial^n g_\gamma(p)}{\partial
  \gamma^n} \ , \hspace{2em} g_\gamma(p) = \log(p\,(\e{\gamma}-1)+1)
\end{equation}
and where we have used
\begin{equation}
  \label{eq:kn-def} k_n = \av{K^{sl}_n(p)} \ .
\end{equation}
The $k_n$ used here refers to a single-locus cumulant; it differs
from $k_n$ used in the main paper, which is the single-locus cumulant
averaged over all loci.  Only when the initial conditions at all loci
are the same will these quantities coincide.

The second observation is a consequence of equations~(\ref{eq:Ksl})
and~(\ref{eq:kn-def}), namely that the cumulants $k_n$ are linearly
related to the allele frequency moments $\mu_n=\av{p^n}$. We can write
the cumulants as
\begin{equation}
  \label{eq:K-def} k_n = \av{K^{sl}_n(p)} = \sum_{l>0} \Kcoef{n}{l}
  \av{p^{l}} = \sum_{l} \Kcoef{n}{l} \mu_l
\end{equation}
where $\Kcoefs{n}{l}$ are the coefficients of the Taylor expansion of
$\av{K^{sl}_n(p)}=g_\gamma(p)$.  They are zero for $l>n$.  We can
write the moments in terms of cumulants
\begin{equation}
  \label{eq:M-def} \mu_l = \sum_{n>0} \Mcoef{l}{n} k_n \ .
\end{equation}
The rest of the calculation involves expanding $K^{sl}_n(p)$ as a
polynomial in $p$ and performing the average to obtain equations in
terms of the moments.  We then expand the moments in terms of the
cumulants $k_n$.  This gives us equations in terms of the coefficients
$\Kcoefs{n}{l}$ and $\Mcoefs{l}{n}$.  We can then use a series of
identities for the coefficient to write the solution in the form of
equation~(\ref{eq:evol}).

\subsubsection*{Identities for Coefficients}

There a six identities among the coefficients that we use
extensively. The first two identities are recursion relations that can
be derived from the generating function given in
equation~(\ref{eq:Ksl})
\begin{eqnarray}
\Kcoef{n}{l} &=& l \Kcoef{n-1}{l} - (l-1) \Kcoef{n-1}{l-1}
\label{eq:Krec} \ , \\ \nonumber \\ \Mcoef{l}{n} &=& \Mcoef{l-1}{n} -
\frac{1}{(l-1)} \Mcoef{l-1}{n-1} \ .
\label{eq:Mrec}
\end{eqnarray}
Using the boundary conditions $\Kcoefs{n}{0} = \Mcoefs{k}{0} = 0$ and
$\Kcoefs{1}{1} = \Mcoefs{1}{1} = 1$ we can generate all the
coefficients.  The next two identities express the fact that the two
sets of coefficients allow us to go from cumulants to moments and back
to cumulants or from moments to cumulants and back to moments
\begin{eqnarray}
\sum_{l>0} \Kcoef{n}{l} \Mcoef{l}{m} &=& \pred{n=m} \label{eq:id1} \ ,
\\
\nonumber \\ \sum_{n>0} \Mcoef{k}{n} \Kcoef{n}{l} &=&
\pred{k=l} \ . \label{eq:id2}
\end{eqnarray}
The final identities relate the coefficients $\Mcoefs{l}{m}$ and
$\Kcoefs{n}{l}$ to binomial coefficients
\begin{eqnarray}
\sum_{l>0} \Kcoef{n}{l+1} \Mcoef{l}{m} &=& -\binom{n-1}{m-1} +
\pred{n=m}
\label{eq:id3} \ , \\ \nonumber \\
\sum_{l>0} l \Kcoef{n}{l} \Mcoef{l}{m} &=& (-1)^{n+m} \binom{n}{m-1} +
\pred{n=m-1} \ .  \label{eq:id4}
\end{eqnarray}
These two identities can be proved by induction using the recursion
relations~(\ref{eq:Krec}) and~(\ref{eq:Mrec}) together with the well
known recursion relation for binomial coefficients.The coefficients used
here are analogous to Stirling numbers in that they obey a second
order recursion relation, although the recursion relations are
different.  Similar identities hold for Stirling numbers; these are
developed in the book Concrete Mathematics (Graham, Knuth and
Patashnik, 2nd Ed. 1995).

\subsubsection*{Derivation of Equation~(\ref{eq:evol})}

We show explicitly how to obtain the terms in
equation~(\ref{eq:evol}) proportional to $\alpha$ and $\beta$, the
remaining (diffusion) term follows similarly, although it requires a
few more lines a algebraic manipulation.  We start from
equation~(\ref{eq:sl-dynamics}).  The term proportional to
$\alpha$ is given by
\begin{eqnarray*}
  \av{\alpha p\,(1-p) \frac{\partial K_n(p)}{\partial p}}
  &=& \av{\alpha p\,(1-p) \sum_{l>0} \Kcoef{n}{l} l p\,^{l-1}} \\ \\
  &=&\alpha \sum_{l>0} \Kcoef{n}{l}\, l \,(\mu_l - \mu_{l+1})\\ \\
  &=&\alpha \sum_{m>0} \sum_{l>0} \Kcoef{n}{l}\, l
  \left(\Mcoef{l}{m}-\Mcoef{l+1}{m} \right) k_m\\ \\
  &=&\alpha \sum_{m>0} \sum_{l>0} \left(l \Kcoef{n}{l} + (l-1)
    \Kcoef{n}{l-1} \right) \Mcoef{l}{m}  k_m \\ \\
  &=& \alpha\sum_{m>0} \sum_{l>0} \Kcoef{n+1}{l}\Mcoef{l}{m} k_m
  = \alpha \, k_{n+1}
\end{eqnarray*}
where we used equations~(\ref{eq:K-def}), (\ref{eq:M-def}), (\ref{eq:Krec})
and~(\ref{eq:id1}) respectively in the derivation.  The term proportional
to $\beta$ is give by
\begin{eqnarray*}
  \av {\beta (1-2p)  \frac{\partial K_n(p)}{\partial p}} &=&
  \av{\beta \sum_{l>0}\Kcoef{n}{l} l (p\,^{l-1}-2p\,^l)} \\ \\
&=& \beta \Kcoef{n}{1} + \beta \sum_{l>0} \left( (l+1) \Kcoef{n}{l+1} -2
  l\Kcoef{n}{l}\right) \av{p\,^l} \\ \\
&=& \beta + \beta\sum_{l>0} \left( (l+1) \Kcoef{n}{l+1} - l\Kcoef{n}{l} -
  l\Kcoef{n}{l} \right) \mu_l \\ \\
&=& \beta + \beta \sum_{l>0} \sum_{m>0} \left( (l+1) \Kcoef{n+1}{l+1} -
  l\Kcoef{n}{l}   \right) \Mcoef{l}{m} k_m \\ \\
&=& \beta + \beta \sum_{m>0} \left( -\binom{n}{m-1} - (-1)^{n+m}
  \binom{n}{m-1} \right) k_m \\ \\
&=& \beta - 2\beta \sum_{m>0} \binom{n}{m-1}\, \pred{\mbox{$n+m$, even}}\, k_m
\end{eqnarray*}
where we used equations~(\ref{eq:Ksl}), (\ref{eq:Krec}), (\ref{eq:M-def}),
(\ref{eq:id3}) and~(\ref{eq:id4}) respectively.  The last term follows
similarly.  Putting these together we obtain equation~(\ref{eq:evol}).

\subsection{Dynamics of the Moments}

Having developed the machinery to go from cumulants to moments we can now
easily obtain a recursion relation for the moments.  Although cumulant
expansions typically have much better convergence properties than moment
expansions (and are therefore more useful in obtaining approximate
solutions), for this problem the dynamics for the moments turns out to have
a simpler form than that for the cumulants.  This allows us to rederive the
diffusion equation for the allele frequencies.

Using equation~(\ref{eq:K-def}) we can write
equation~(\ref{eq:sl-dynamics}) in terms of moments
\begin{equation}
  \label{eq:moment-eqn}
   \sum_{l>0} \Kcoef{n}{l} \frac{\dd \mu_l}{\dd t} = \sum_{l>0}
    \Kcoef{n}{l} l \left( \alpha (\mu_l - \mu_{l+1}) + \beta
    (\mu_{l-1}-2\mu_l) + \frac{(l-1)}{2} (\mu_{l-1}-\mu_{l}) \strut
    \right) \ .
\end{equation}
We can obtain a differential equation for the moments by multiplying both
sides of equation~(\ref{eq:moment-eqn}) by $\Mcoefs{k+1}{n}$ and summing
over $n$.  Using the identity~(\ref{eq:id2}) we find
\begin{equation}
  \label{eq:moments-diff1}
  \frac{\dd \mu_{k+1}}{\dd t} = (k+1)\left(\alpha (\mu_{k+1}-\mu_{k+2}) +
  \beta(\mu_k-2 \mu_{k+1}) +  \frac{k}{2}(\mu_k-\mu_{k+1}) \strut
  \right) \ .
\end{equation}
Rearranging gives equation~(\ref{eq:moments-diff}).

Equation~(\ref{eq:moments-diff}) is a linear transformation of the
differential equation~(\ref{eq:evol}) for the cumulants.  To see
this more directly we can consider the coefficients
$U_{l,n}=\Mcoefs{l}{n}$ as elements of a lower-triangular matrix
$\bm{U}$.  The coefficients $\Kcoefs{n}{l}$ are the elements of
the matrix $\bm{U}^{-1}$.  We can write
equation~(\ref{eq:moments-diff}) as a matrix equation analogous to
equation~(12)
\begin{equation}
  \label{eq:moment-matrix-eqn}
  \frac{\dd \bm{\mu}}{\dd t} = \bm{U} \frac{\dd \bm{k}}{\dd t}
  = - \bar{\bm{M}} \bm{\mu} + \bar{\bm{d}}
\end{equation}
where $\bar{\bm{M}}=\bm{U} \bm{M} \bm{U}^{-1}$ and
$\bar{\bm{d}}=\bm{U}\bm{d}$ with $d_k=\beta\pred{k=1}$.  The matrix
$\bar{\bm{M}}$ is tridiagonal.

\subsubsection*{Obtaining the Diffusion Equation}

Using equation~(\ref{eq:moments-diff}), we can obtain the diffusion
equation for allele frequencies.  We first obtain a partial differential
equation for the characteristic function
\begin{displaymath}
  \tilde{\phi}(z, t) = 1+\sum_{k>0} \frac{\mu_k\, (\ii z)^k}{k!} = \int
  \e{\, \ii z p} \, \phi(p,t) \,\dd p \ .
\end{displaymath}
To do this, we multiplying equation~(\ref{eq:moments-diff})
by $(\ii z)^{k+1}/k!$ and sum over $k$, giving
\begin{equation}
  \label{eq:pde}
  \frac{\partial \tilde{\phi}}{\partial t} = - \ii z \left( -\left(\alpha +
  \frac{\ii z}{2}\right) \frac{\partial^2 \tilde{\phi}}{\partial z^2} +
  \ii \left(2\beta-\alpha-\frac{\ii z}{2}\right)  \frac{\partial
  \tilde{\phi}}{\partial z}   - \beta \,\tilde{\phi}\right) \ .
\end{equation}
Taking the inverse Fourier transform we obtain the diffusion
equation for the single-locus allele frequencies
\begin{equation}
  \label{eq:diffusion-eqn}
  \frac{\partial \phi}{\partial t} = \frac{\partial \ }{\partial p} \left(
    \frac{p\,(1-p)}{2} \frac{\partial \phi}{\partial p} - \left(\alpha
    p\,(1-p) + \left(\beta-\frac{1}{2}\right) (1-2p) \right) \phi
  \right) \ .
\end{equation}
This can be re-written in the form of equation~(4).  Setting the right-hand
side of this equation to zero we easily obtain the fixed point solution
$\phi(p,\infty)$, which was first found by Wright~(1937).  Since the
diffusion equation was derived directly from the moment
equations~(\ref{eq:moments-diff}), the fixed points will have the same
moments, and as equations~(\ref{eq:moments-diff}) are just a linear
transformation of cumulant equations~(\ref{eq:evol}), the fixed point will
have the same cumulants.

We can obtain eigenvalue equations from the diffusion equation by looking
for solutions of the form $\phi(p,t) = \phi(p,\infty)+\rho(t) \psi(p)$.
This reduces the problem to solving a eigenvalue differential equation,
subject to boundary conditions.  However, these equations are, in general,
difficult to solve as they require solving a boundary value problem,
often with complicated boundary conditions.

The similarity between diffusion equation problem and quantum mechanics has
often been observed.  Here we note another connection.  Moving between
a matrix equation and diffusion equation parallels the connection
between Heisenberg's matrix formulation of quantum mechanics and
Schr\"odinger's differential equation.

\newpage


\begin{figure}[h]
\caption{The first two cumulants are shown from a single
realisation of the evolutionary dynamics for populations with
differing numbers of loci: $L=32$ (left) and $L=3200$ (right). Both populations evolve under
multiplicative selection, reversible mutation and free recombination, with $N=500$,
$\alpha=5$ and $\beta=0.01$ ($\alpha\equiv Ns$ and $\beta\equiv Nu$).}
\label{fig:fluc}
\end{figure}

\begin{figure}[h]
\caption{Fixed point cumulant results from the truncated cumulant
equations ($k_n^*$) are compared to cumulants derived from Wright's
distribution ($\tilde{k_n}$) as the truncation order is increased. The
top figures show the difference between the first and second
cumulant results for $\beta=1$, using a logarithmic scale. The bottom
figures show the first cumulant results for lower mutation rates,
$\beta=0.1$ and $\beta=10^{-3}$. The symbols denote different selection intensities, with
$\alpha=1(+)$, $2(\triangle)$, $5(\mbox{o})$ and $10(\times)$.}
\label{fig:fp}
\end{figure}

\begin{figure}[h]
\caption{The averaged dynamics are shown for the first two cumulants with
$\alpha=1(\mbox{o})$, $5(\triangle)$ and $10(\times)$. The eight cumulant
theory is shown by the solid lines. The results were averaged
over $500$ iterations of the evolutionary dynamics with free
recombination, $N=200$ and $\beta=0.01$. In each case the population was initialised
with all alleles set to zero.}
\label{fig:dynamics}
\end{figure}

\begin{figure}[h]
\caption{Approximate results for the two lowest eigenvalues given by the
perturbation theory are compared to results from the 16 cumulant
truncated system (solid lines). The 2nd order (dot-dashed) and 6th
order (dashed) perturbation theory results are shown for various $\alpha$,
for $\beta=0.1$ (left) and $\beta=0.01$ (right).}
\label{fig:eval}
\end{figure}

\begin{figure}[h]
\caption{The average rate of change in the mean is shown as a function
of the mean for $\alpha=1$ (left) and $\alpha=10$ (right). The symbols show
results averaged over up to $1000$ iterations of the evolutionary
dynamics, with $\beta=1(\mbox{o})$, $0.1(\times)$ and
$0.01(\triangle)$. The solid line shows the eight cumulant adiabatic
elimination result for small $\beta$ (see Eqs.~(\ref{eq:dk1dt}) and
(\ref{eq:k2qe})). Populations were either initialised with alleles all set to
zero, giving a positive rate of change, or were initialised
with alleles all set to one, giving a negative rate of change. The
populations were then given enough time to converge close to the fixed
point. For $\alpha=1$ we chose $L=960$ and $N=100$ and for $\alpha=10$ we
chose $L=96$ and $N=200$. In each case the populations were subject
to free recombination.}
\label{fig:pop_vel}
\end{figure}

\newpage

\pagestyle{empty}

\begin{figure}[b]
	\setlength{\unitlength}{1.0cm}
	\begin{center}
	\begin{picture}(15,15)
	\epsfysize = 9cm
	\put(-0.5,0){\epsfbox[50 0 550 600]{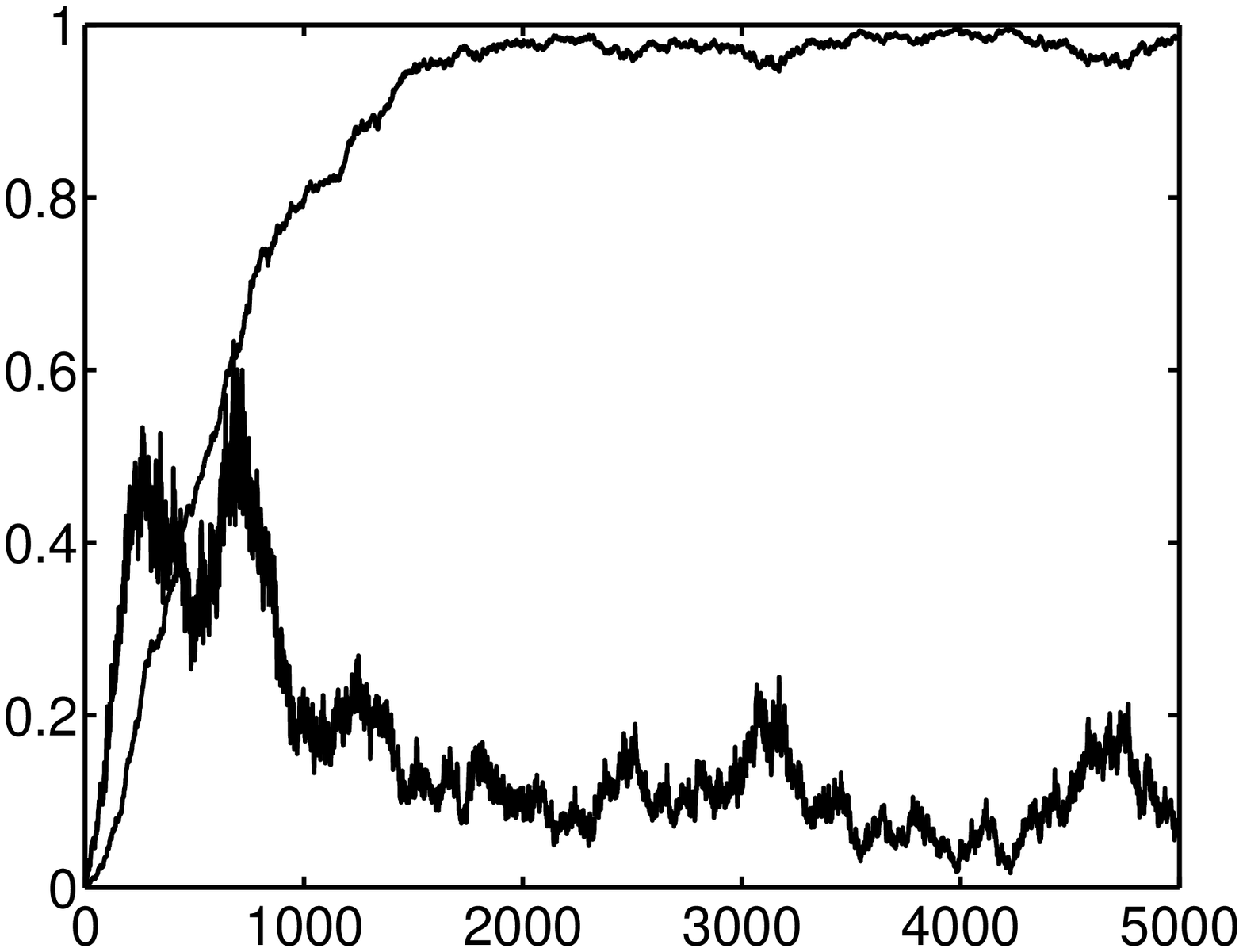}}
	\epsfysize = 9cm
	\put(9,0){\epsfbox[50 0 550 600]{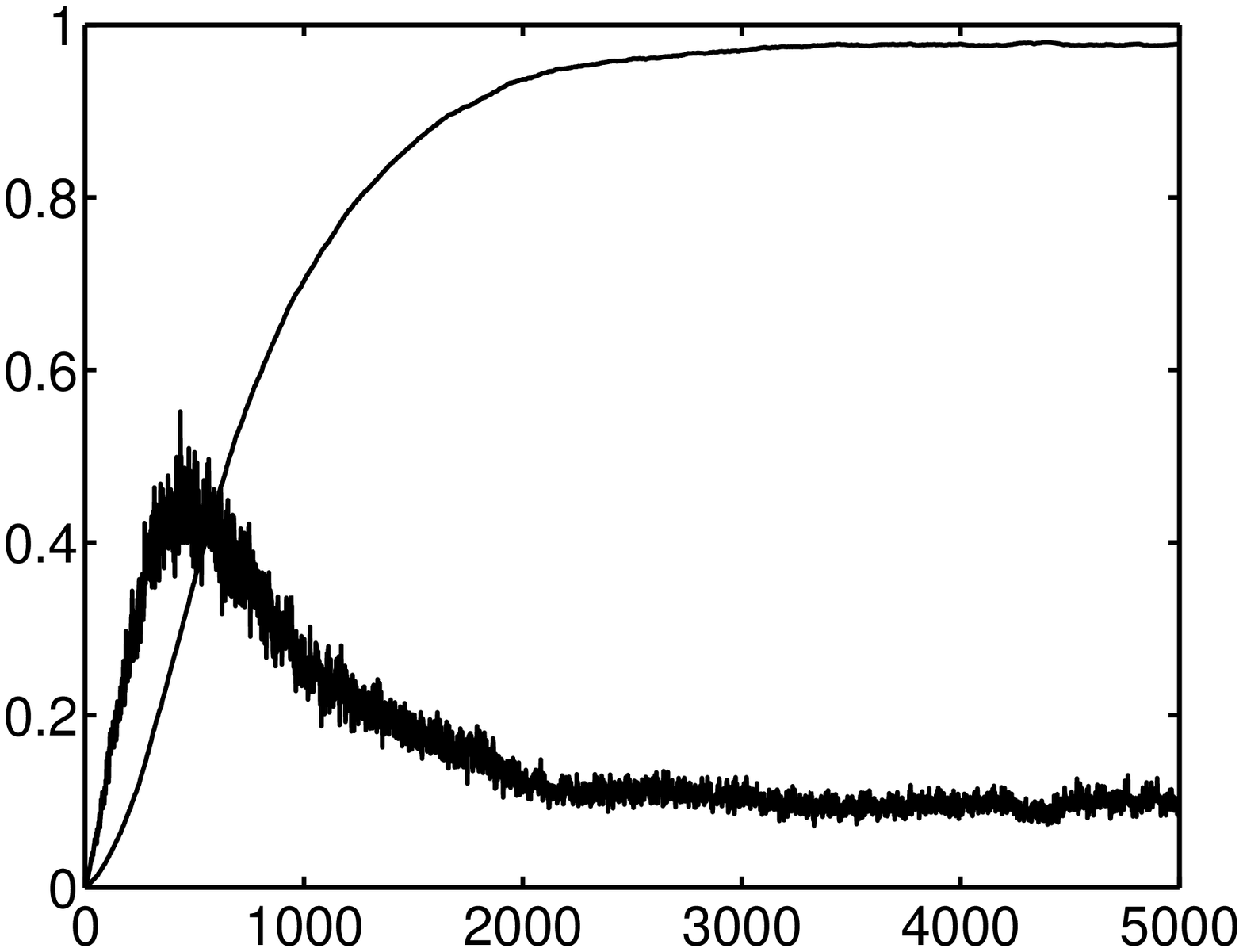}}
	\put(2.2,2.3){\Large{Generation}}
	\put(2.8,9){\large{\mbox{$L=32$}}}
	\put(2.4,7.8){\large{\mbox{$K_1/L$}}}
	\put(2.4,4.7){\large{\mbox{$5K_2/L$}}}
	\put(11.8,2.3){\Large{Generation}}
	\put(12,9){\large{\mbox{$L=3200$}}}
	\put(12,7.8){\large{\mbox{$K_1/L$}}}
	\put(12,4.7){\large{\mbox{$5K_2/L$}}}
	\put(7.5,-2.0){\LARGE{Fig. 1}}
	\end{picture}
	\end{center}
\end{figure}

\newpage

\begin{figure}[b]
	\setlength{\unitlength}{1.0cm}
	\begin{center}
	\begin{picture}(20,20)
	\epsfysize = 9cm
	\put(0,9){\epsfbox[50 0 550 600]{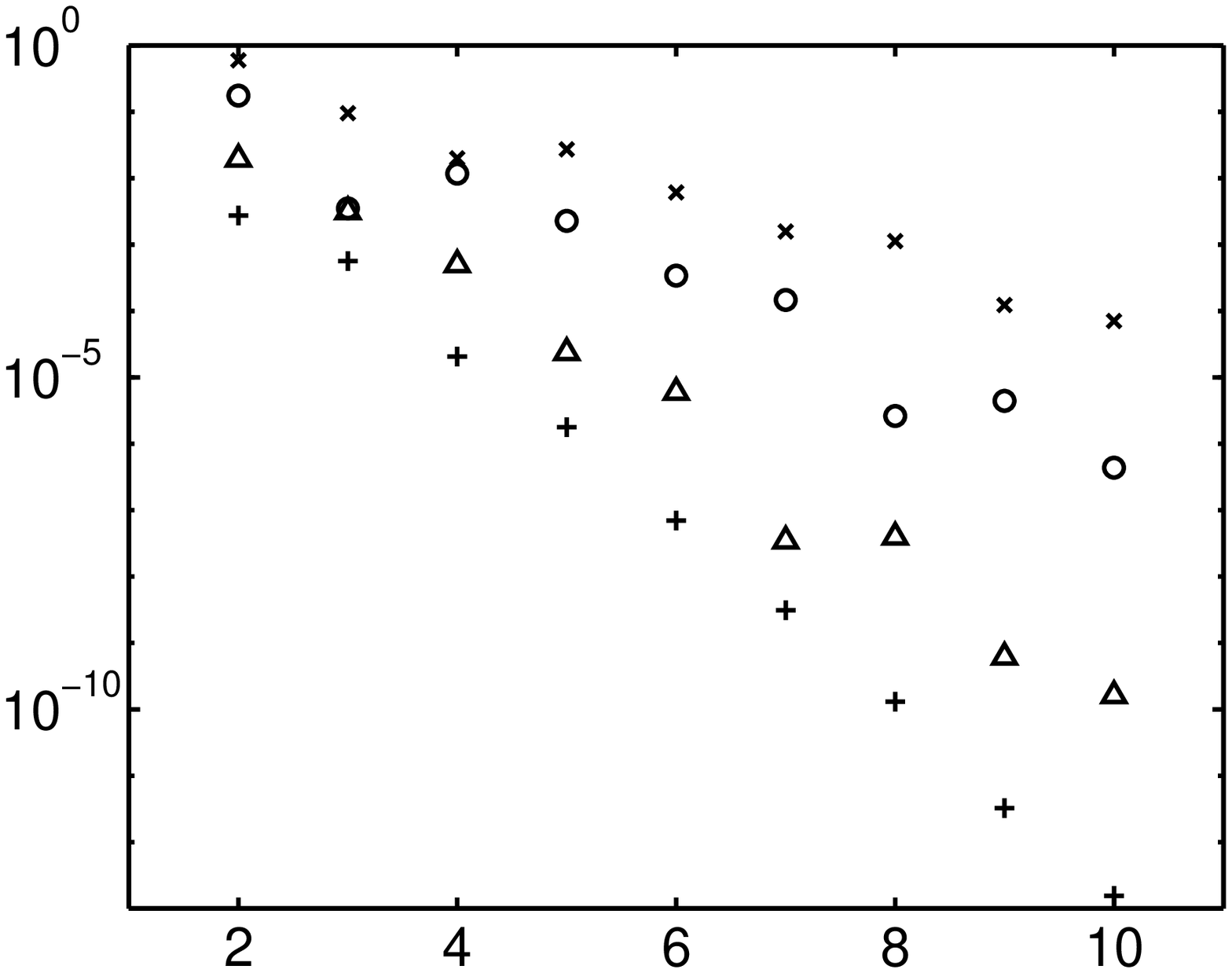}}
	\epsfysize = 9cm
	\put(9,9){\epsfbox[50 0 550 600]{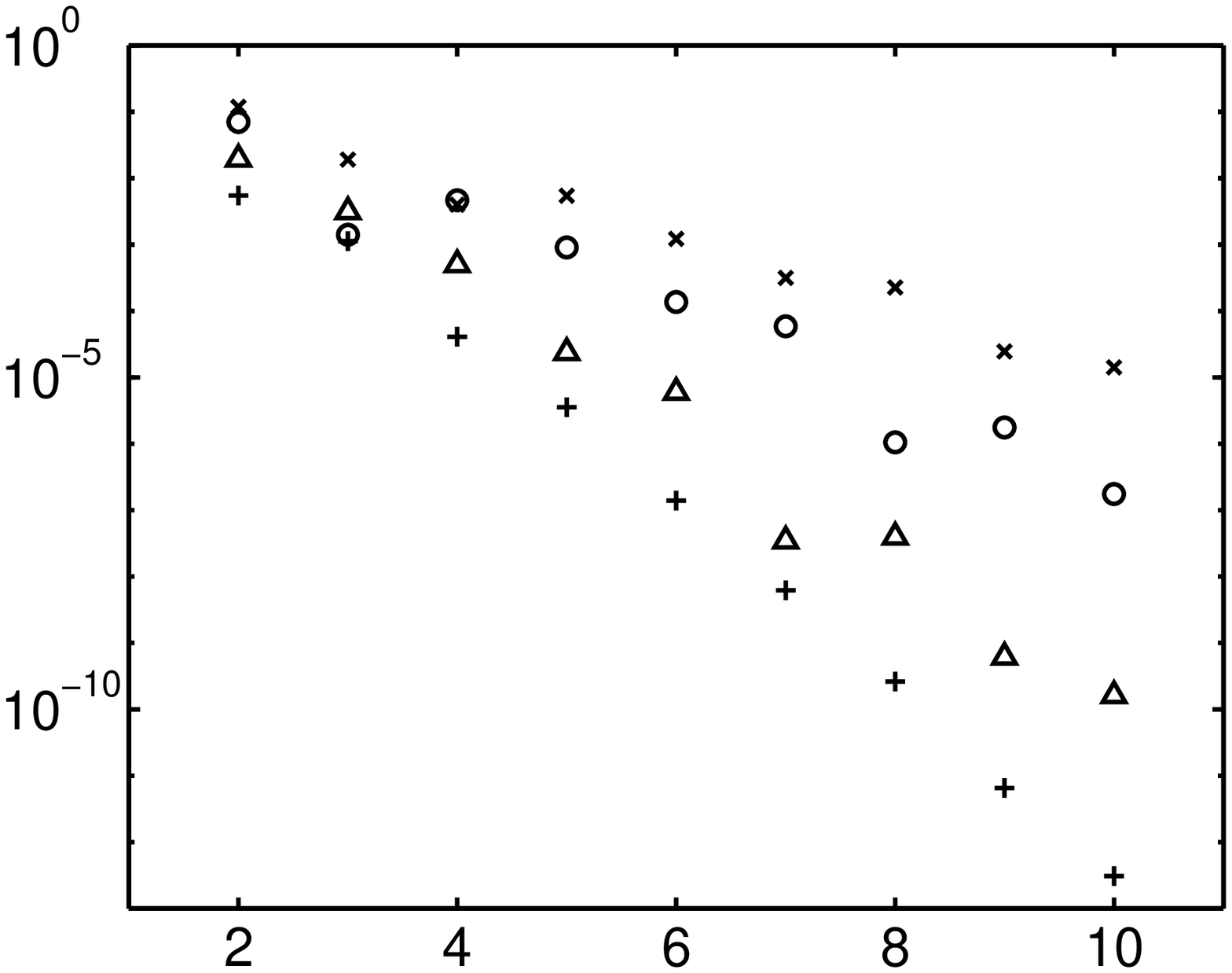}}
	\put(3.5,18){\large{\mbox{$\beta=1$}}}
	\put(-1.3,16.5){\large{\mbox{$|k_1^*-\tilde{k}_1|$}}}
	\put(2.6,11.4){\large{No. of cumulants}}
	\put(12.5,18){\large{\mbox{$\beta=1$}}}
	\put(7.7,16.5){\large{\mbox{$|k_2^*-\tilde{k}_2|$}}}
	\put(11.4,11.4){\large{No. of cumulants}}
	\epsfysize = 9cm
	\put(0,1){\epsfbox[50 0 550 600]{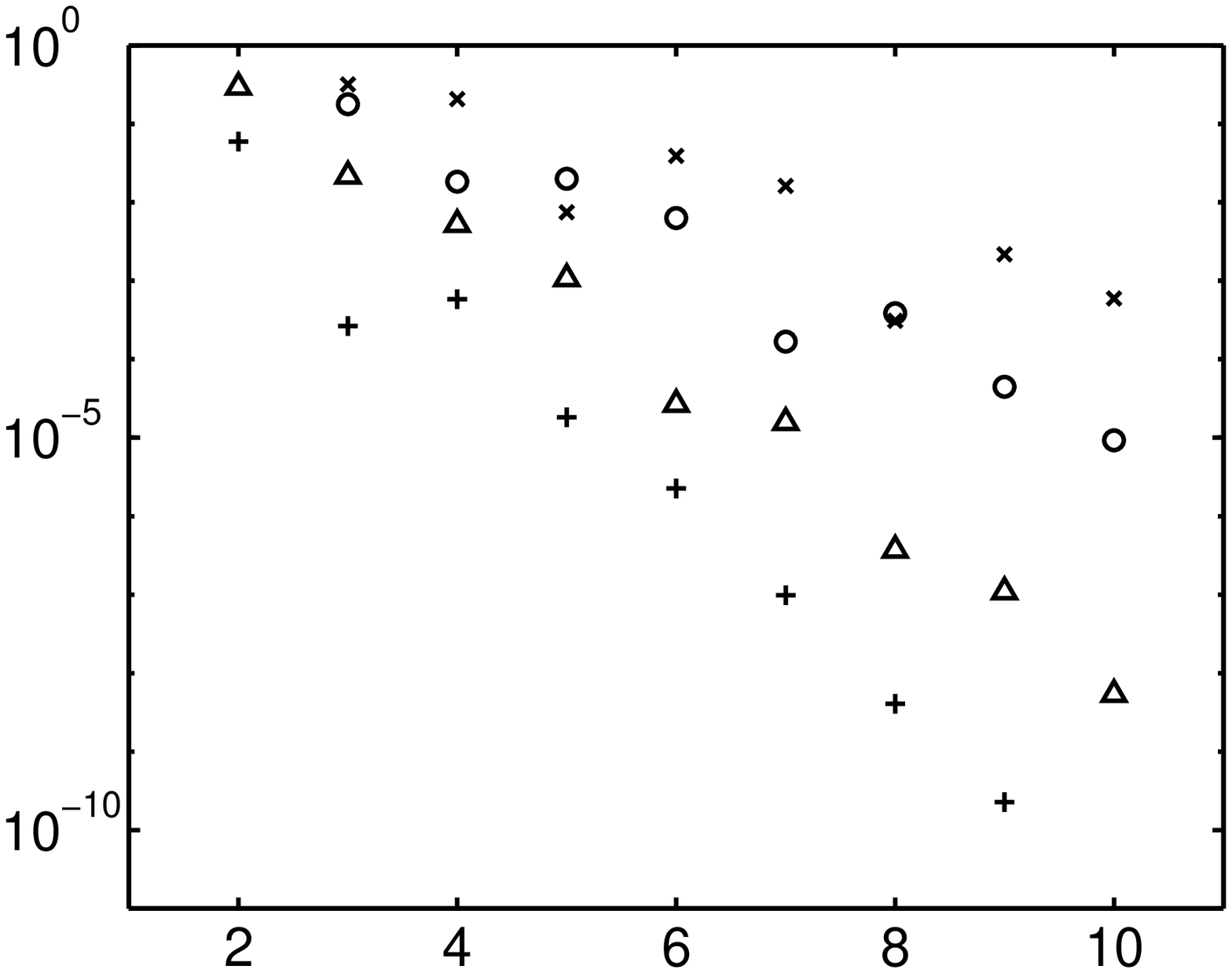}}
	\epsfysize = 9cm
	\put(9,1){\epsfbox[50 0 550 600]{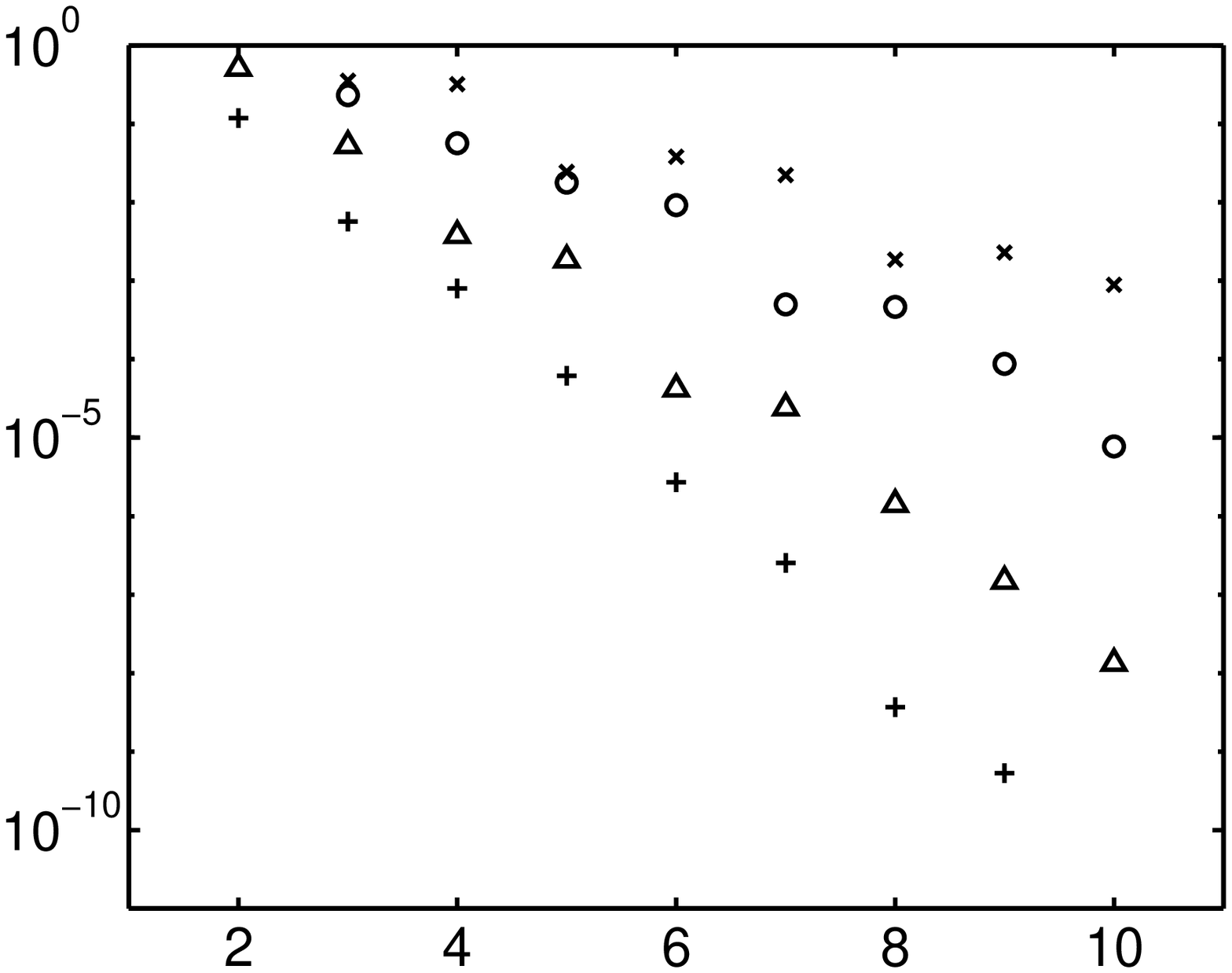}}
	\put(3.5,10){\large{\mbox{$\beta=0.1$}}}
	\put(-1.3,8.5){\large{\mbox{$|k_1^*-\tilde{k}_1|$}}}
	\put(2.6,3.4){\large{No. of cumulants}}
	\put(12.5,10){\large{\mbox{$\beta=10^{-3}$}}}
	\put(7.7,8.5){\large{\mbox{$|k_1^*-\tilde{k}_1|$}}}
	\put(11.4,3.4){\large{No. of cumulants}}
	\put(7.5,-1.0){\LARGE{Fig. 2}}
	\end{picture}
	\end{center}
\end{figure}

\newpage

\begin{figure}[b]
	\setlength{\unitlength}{1.0cm}
	\begin{center}
	\begin{picture}(15,15)
	\epsfysize = 9cm
	\put(-0.5,0){\epsfbox[50 0 550 600]{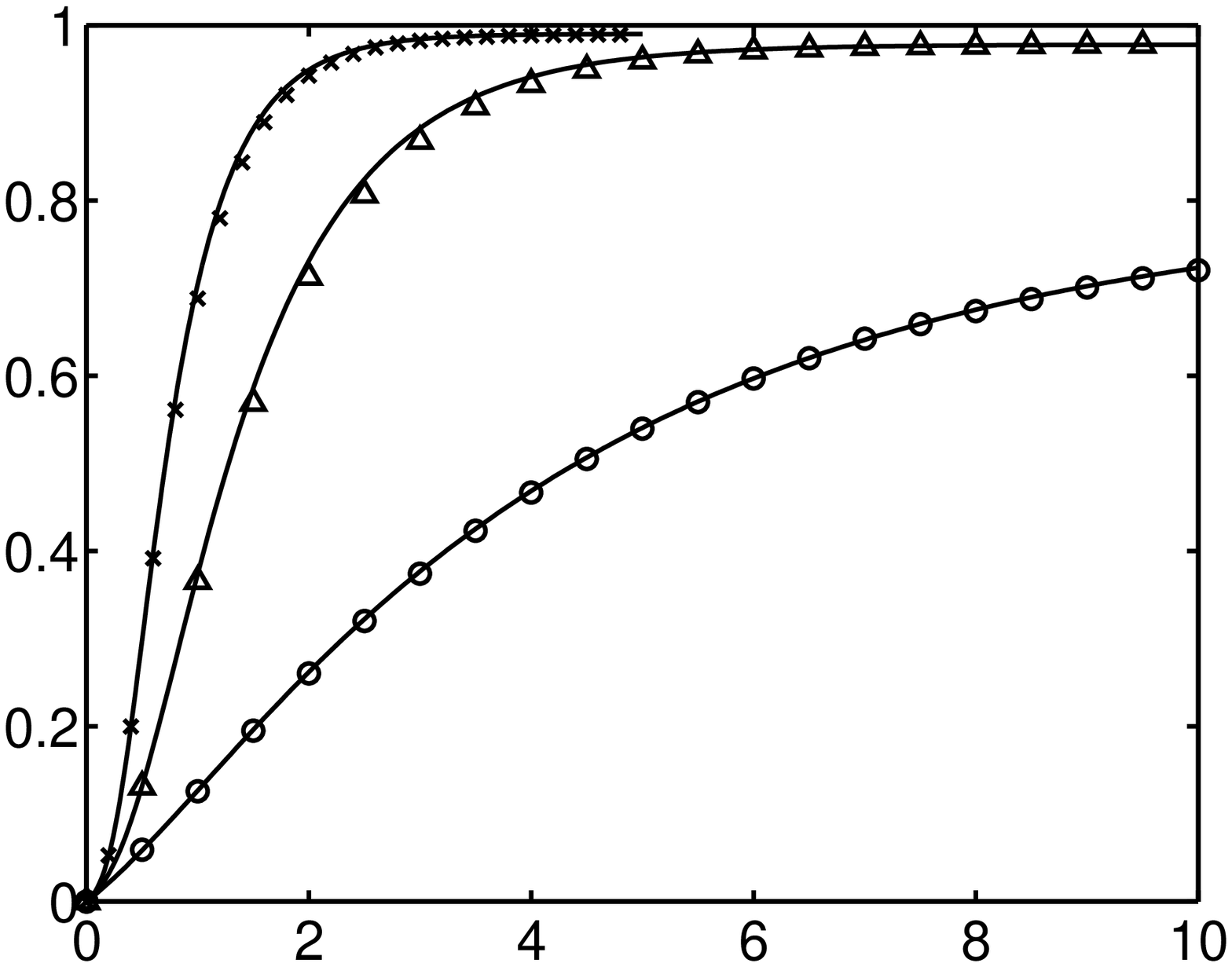}}
	\epsfysize = 9cm
	\put(9,0){\epsfbox[50 0 550 600]{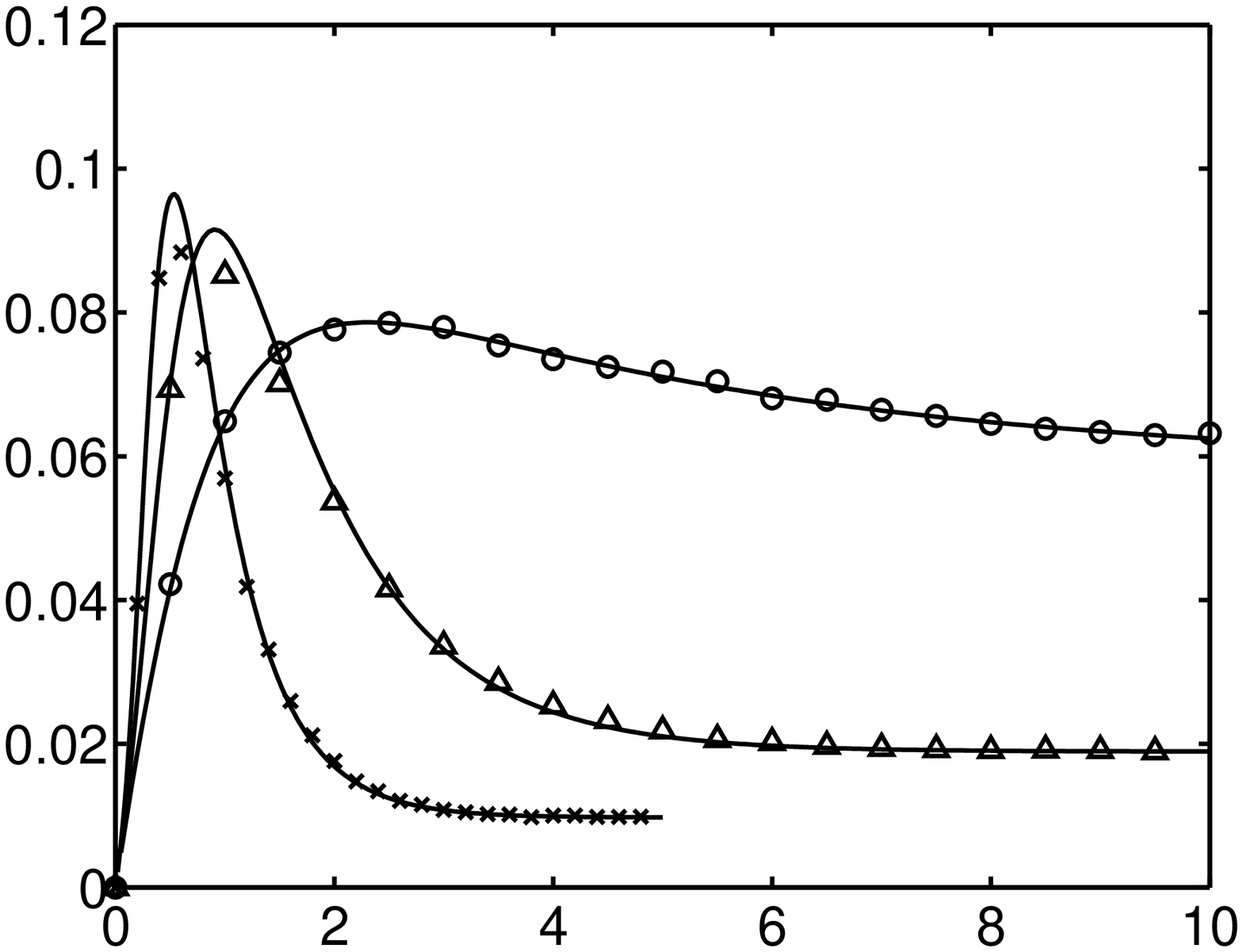}}
	\put(-1,7.0){\Large{\mbox{$k_1$}}}
	\put(3.3,2.4){\Large{\mbox{$t$}}}
	\put(8.2,7.0){\Large{\mbox{$k_2$}}}
	\put(12.9,2.4){\Large{\mbox{$t$}}}
	\put(6.5,-2.0){\LARGE{Fig. 3}}
	\end{picture}
	\end{center}
\end{figure}

\newpage

\begin{figure}[b]
	\setlength{\unitlength}{1.0cm}
	\begin{center}
	\begin{picture}(20,20)
	\epsfysize = 9cm
	\put(0,9){\epsfbox[50 0 550 600]{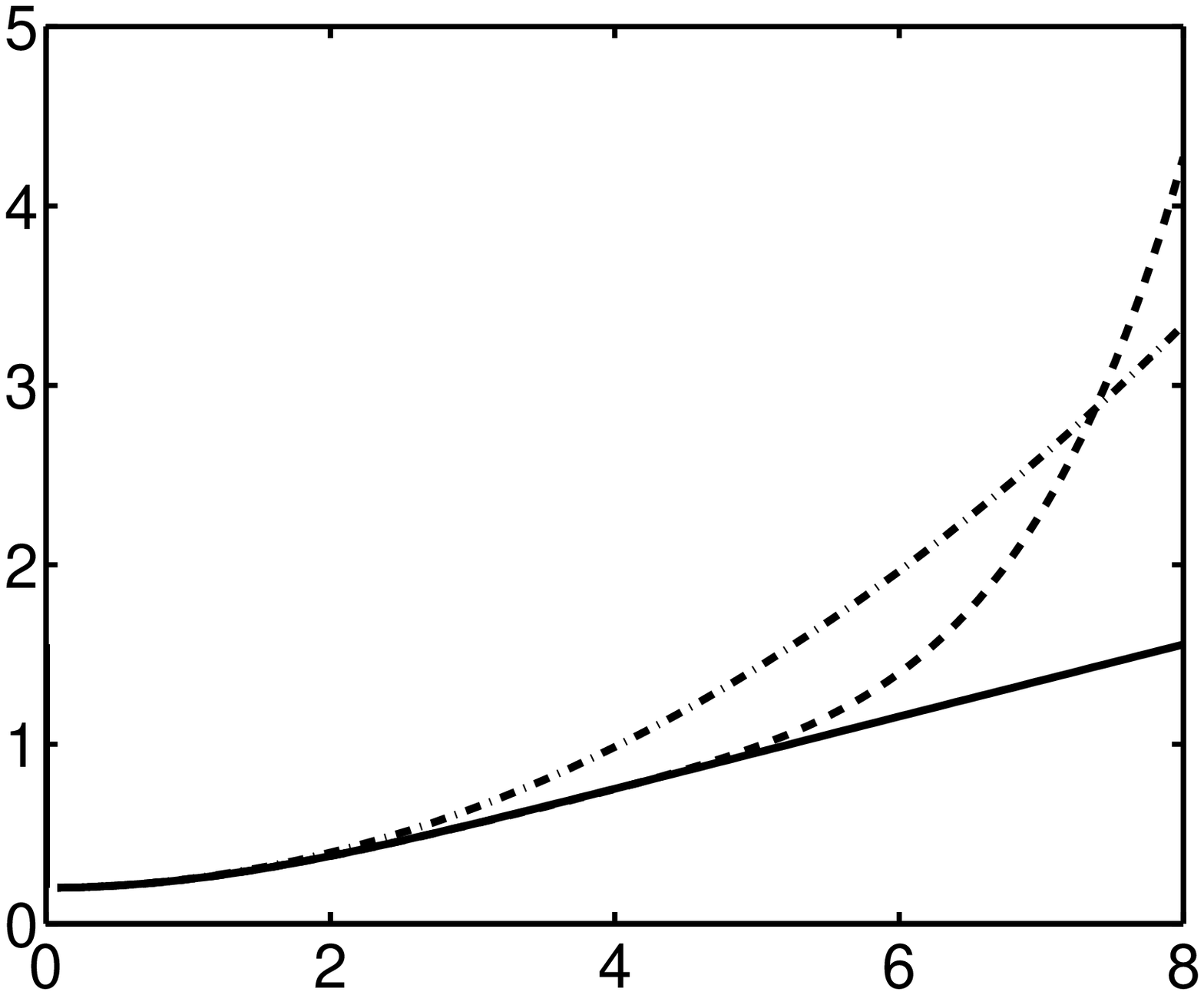}}
	\epsfysize = 9cm
	\put(9,9){\epsfbox[50 0 550 600]{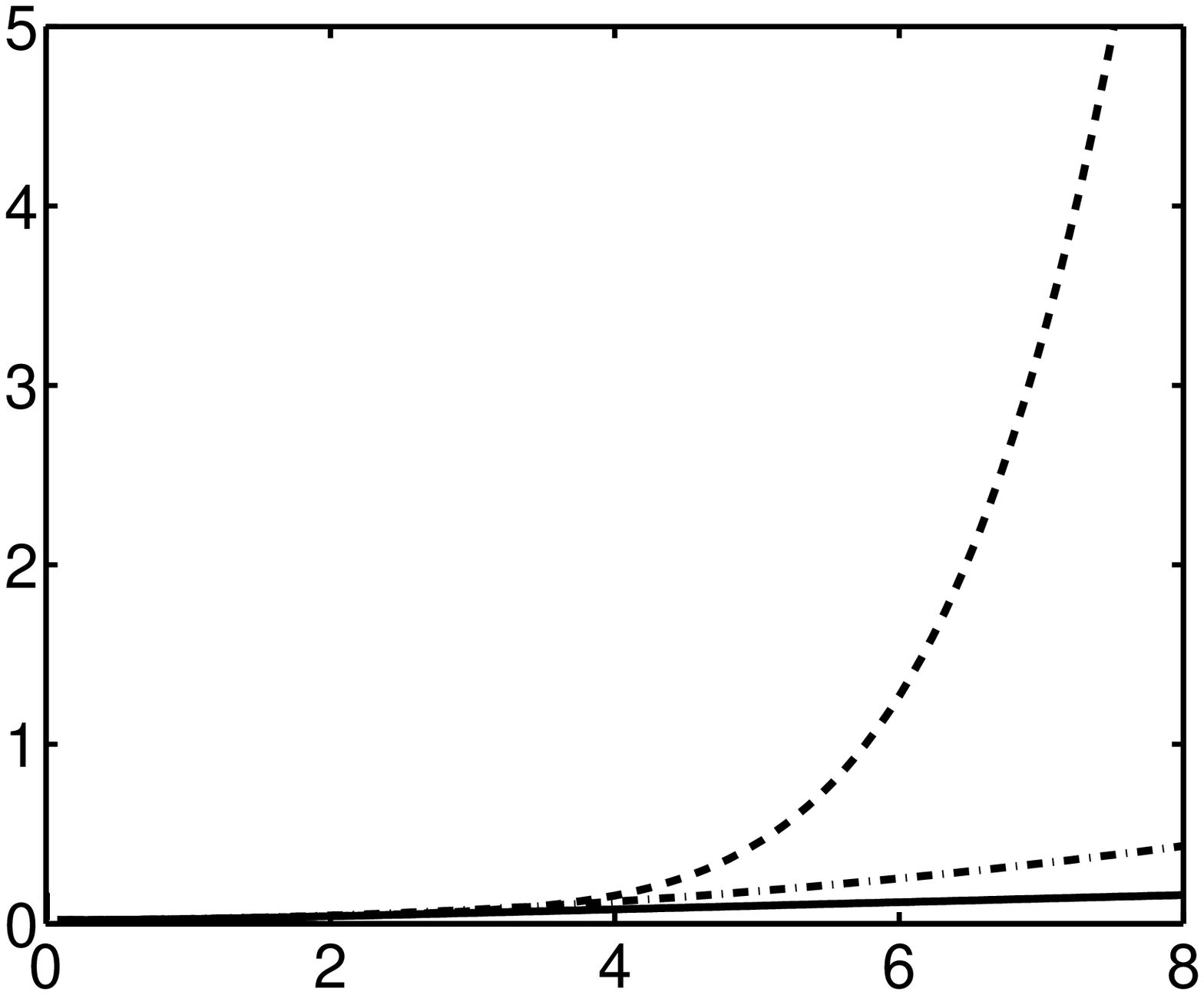}}
	\put(3.1,18){\large{\mbox{$\beta=0.1$}}}
	\put(-0.6,16.5){\Large{\mbox{$\lambda_1$}}}
	\put(3.6,11.4){\large{$\alpha$}}
	\put(12.1,18){\large{\mbox{$\beta=0.01$}}}
	\put(8.4,16.5){\Large{\mbox{$\lambda_1$}}}
	\put(12.6,11.4){\large{$\alpha$}}
	\epsfysize = 9cm
	\put(0,1){\epsfbox[50 0 550 600]{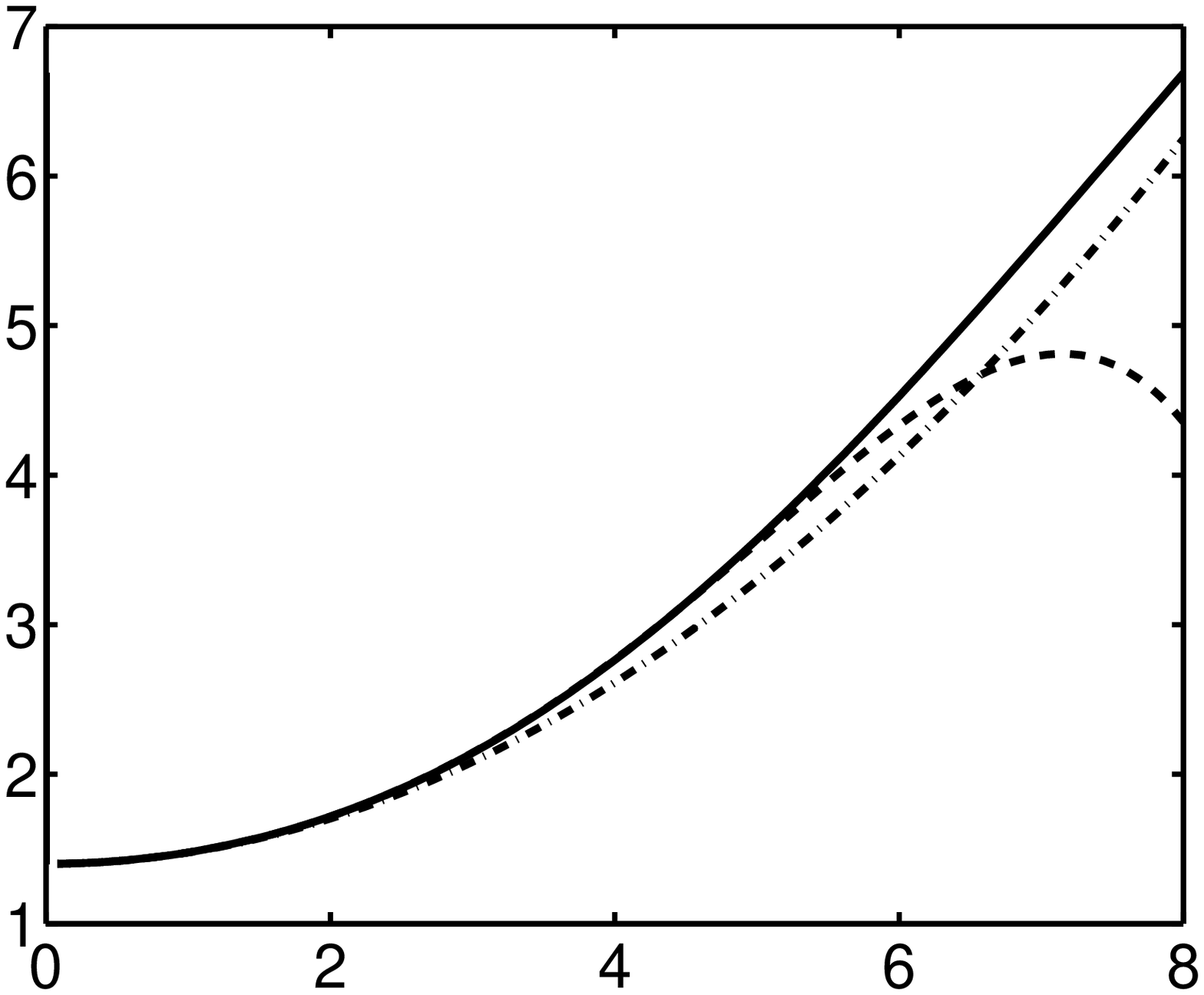}}
	\epsfysize = 9cm
	\put(9,1){\epsfbox[50 0 550 600]{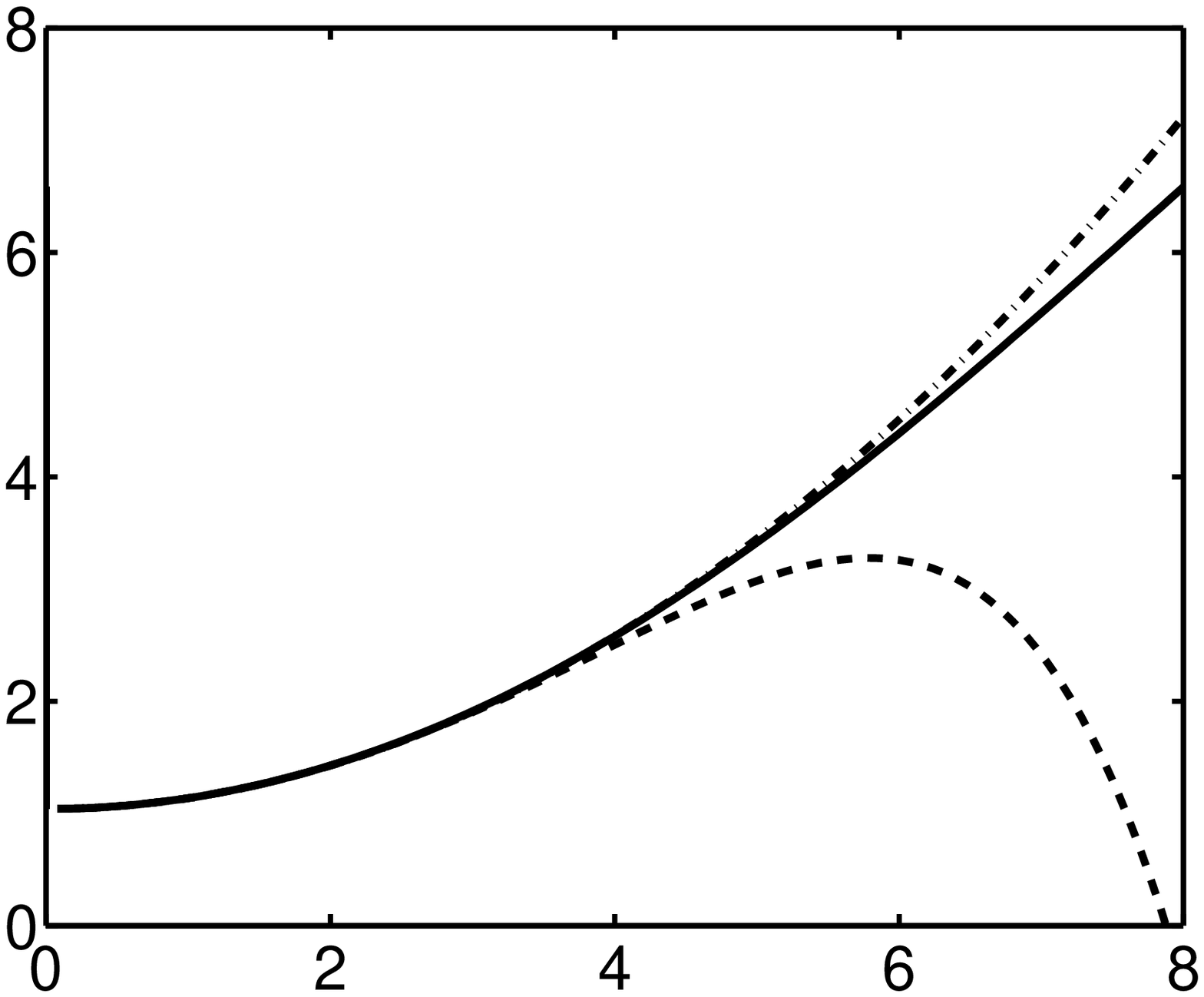}}
	\put(3.1,10){\large{\mbox{$\beta=0.1$}}}
	\put(-0.6,8.5){\Large{\mbox{$\lambda_2$}}}
	\put(3.6,3.4){\large{$\alpha$}}
	\put(12.1,10){\large{\mbox{$\beta=0.01$}}}
	\put(8.4,8.5){\Large{\mbox{$\lambda_2$}}}
	\put(12.6,3.4){\large{$\alpha$}}
	\put(7.5,-1.0){\LARGE{Fig. 4}}
	\end{picture}
	\end{center}
\end{figure}

\newpage

\begin{figure}[b]
	\setlength{\unitlength}{1.0cm}
	\begin{center}
	\begin{picture}(15,15)
	\epsfysize = 9cm
	\put(-0.5,0){\epsfbox[50 0 550 600]{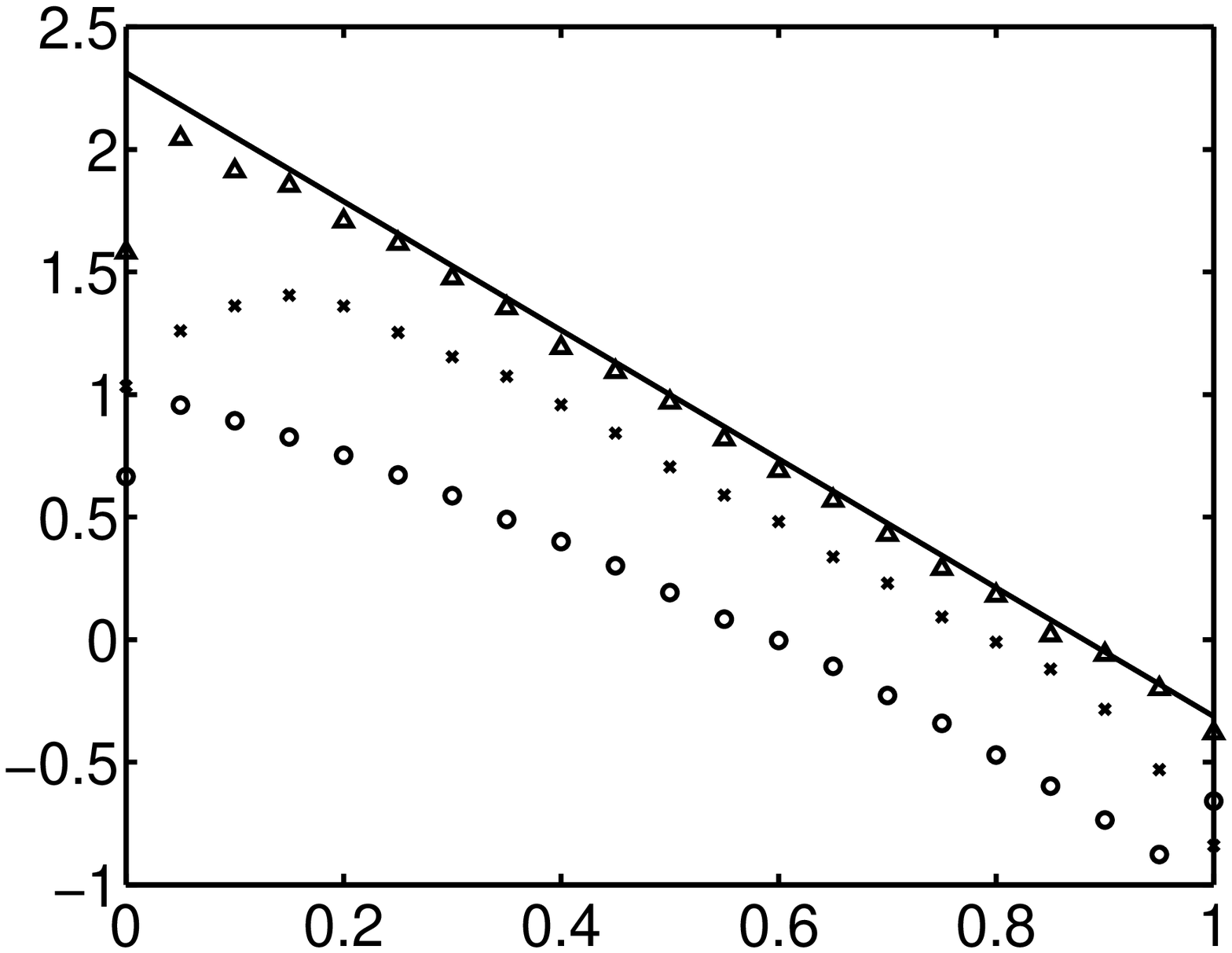}}
	\epsfysize = 9cm
	\put(9,0){\epsfbox[50 0 550 600]{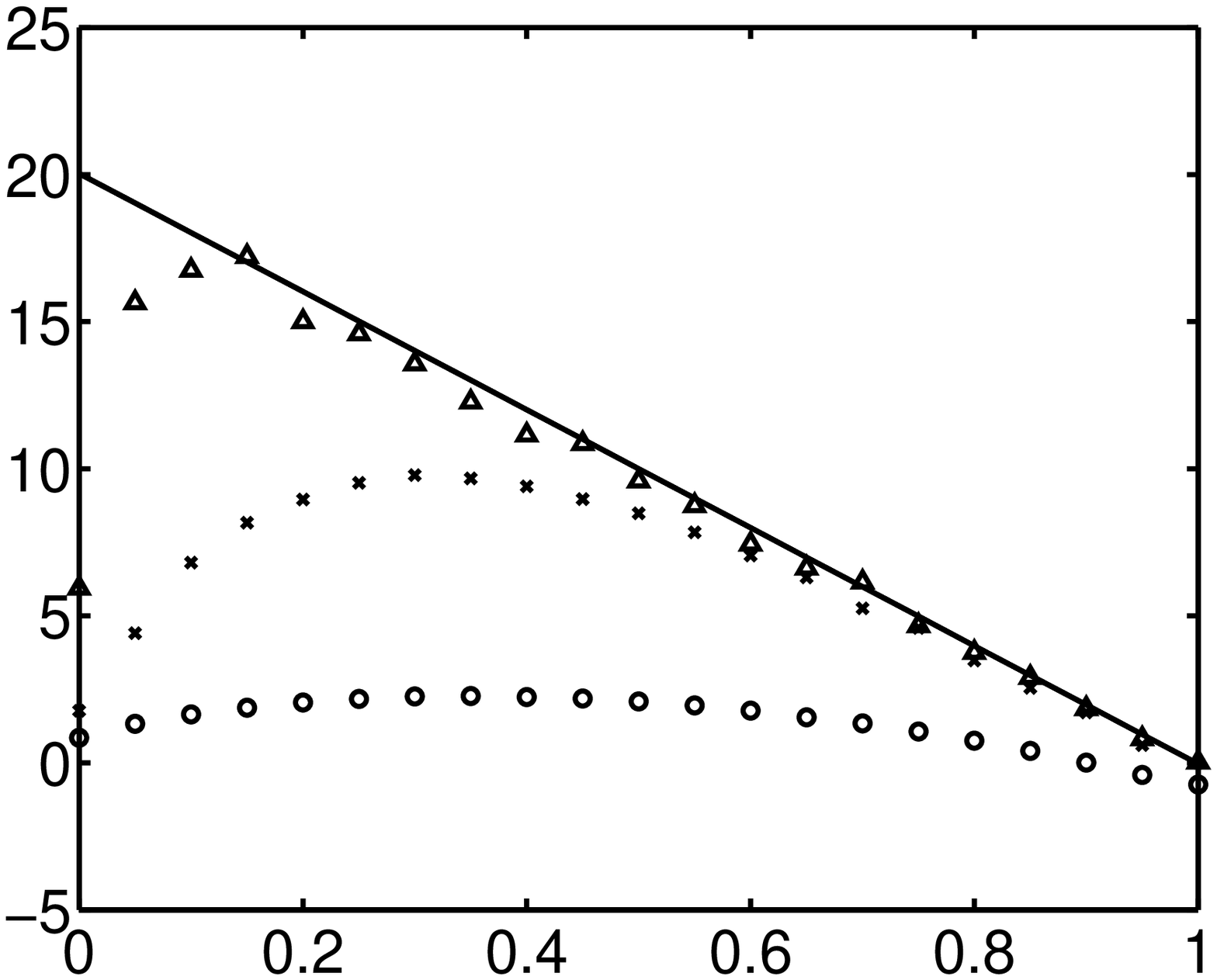}}
	\put(-2.0,7.0){\large{\mbox{$\displaystyle{\frac{1}{\beta}\frac{\mathrm{d}k_1}{\mathrm{d}t}}$}}}
	\put(3.3,2.2){\Large{\mbox{$k_1$}}}
	\put(2.8,9){\large{\mbox{$\alpha=1$}}}
	\put(7.6,7.0){\large{\mbox{$\displaystyle{\frac{1}{\beta}\frac{\mathrm{d}k_1}{\mathrm{d}t}}$}}}
	\put(12.9,2.2){\Large{\mbox{$k_1$}}}
	\put(12,9){\large{\mbox{$\alpha=10$}}}
	\put(6.5,-2.0){\LARGE{Fig. 5}}
	\end{picture}
	\end{center}
\end{figure}


\end{document}